\documentclass[twocolappendix]{emulateapj}
\usepackage{subfigure}
\usepackage{amsmath}

\newcommand{\bmath}[1]{\mbox{\boldmath{$#1$}}}

\newcommand{\del}{{\bf \nabla}}
\newcommand{\alf}{{\rm Alfv\acute{e}n}}
\newcommand{\cs}{c_{\rm s}}

\begin{document}

\title{Turbulence in the Outer Regions of Protoplanetary Disks. \\ II. Strong Accretion Driven by a Vertical Magnetic Field}

\author{Jacob B. Simon\altaffilmark{1}, Xue-Ning Bai\altaffilmark{2,3}, Philip J. Armitage\altaffilmark{1,5},
James M. Stone\altaffilmark{4}, and Kris Beckwith\altaffilmark{1,6}}

\email{jbsimon@jila.colorado.edu}

\begin{abstract}
\noindent
We carry out a series of local, vertically stratified shearing box simulations of protoplanetary disks that include ambipolar diffusion and a net vertical magnetic field.  The ambipolar diffusion profiles we employ correspond to 30AU and 100AU in a minimum mass solar nebula (MMSN) disk model, which consists of a far-UV-ionized surface layer and low-ionization disk interior. These simulations serve as a follow up to \cite{simon13a}, in which we found that without a net vertical field, the turbulent stresses that result from the magnetorotational instability (MRI) are too weak to account for observed accretion rates.  The simulations in this work show a very strong dependence of the accretion stresses on the strength of the background vertical field; as the field strength increases, the stress amplitude increases.  For a net vertical field strength (quantified by $\beta_0$, the ratio of gas to magnetic pressure at the disk mid-plane) of $\beta_0 = 10^4$ and $\beta_0 = 10^5$, we find accretion rates $\dot{M} \sim 10^{-8}$--$10^{-7} M_{\sun}/{\rm yr}$.  These accretion rates agree with observational constraints, suggesting a vertical magnetic field strength of $\sim 60$--200 $\mu$G and 10--30~$\mu$G at 30 AU and 100 AU, respectively, in a MMSN disk. Furthermore, the stress has a non-negligible component due to a magnetic wind.   For sufficiently strong vertical field strengths, MRI turbulence is quenched, and the flow becomes largely laminar, with accretion proceeding through large scale correlations in the radial and toroidal field components as well as through the magnetic wind.  In all simulations, the presence of a low ionization region near the disk mid-plane, which we call the ambipolar damping zone, results in reduced stresses there. 
\end{abstract} 


\altaffiltext{1}{JILA, University of Colorado and NIST, 440 UCB, Boulder, CO 80309-0440}
\altaffiltext{2}{Institute for Theory and Computation, Harvard-Smithsonian Center for Astrophysics,
60 Garden St., MS-51, Cambridge, MA 02138}
\altaffiltext{3}{Hubble Fellow}
\altaffiltext{4}{Department of Astrophysical Sciences, Princeton University, Princeton, NJ 08544}
\altaffiltext{5}{Department of Astrophysical and Planetary Sciences, University of Colorado, Boulder, CO 80309}
\altaffiltext{6}{Tech-X Corporation, 5621 Arapahoe Ave., Suite A, Boulder, CO 80303}

\section{Introduction}

The structure and evolution of the gas and dust in protoplanetary disks play an integral role in the
formation of stars and planets.  Turbulent angular momentum transport \citep{shakura73} allows disk gas to
accrete onto the central protostar and determines the global density
distribution within which planets will form.  Turbulence also influences the earliest stages of planet
formation by inhibiting or enhancing dust coagulation and settling towards the disk mid-plane
\cite[e.g.,][]{dubrulle95, ormel07,youdin07,birnstiel11}.

Despite its importance, the nature of turbulence in protoplanetary disks is not well constrained.  The
magnetorotational instability \cite[MRI;][]{balbus98} generates turbulence, but the efficiency of the MRI
depends (often times, quite strongly) on the ionization fraction of the gas. Apart from regions very close
to the star, protoplanetary disks are too cold to be thermally ionized and the coupling to magnetic fields
is instead dependent on external sources of ionization. Across large regions of the disk the predicted
ionization level is very low, motivating disk models \citep{gammie96} in which MRI-driven turbulence is
substantially suppressed or quenched entirely \citep[e.g.,][]{ilgner06,bai11b,perez-becker11b,mohanty13}.  
The MRI can also be suppressed in regions of high ionization, either due to poor coupling between the field 
and fluid at low densities and high magnetic field strengths \citep{bai11a} or when the MRI stability condition is satisfied \citep{balbus98}.

Three non-ideal magnetohydrodynamic (MHD) effects are important in weakly ionized protoplanetary disks:
Ohmic resistivity, the Hall effect, and ambipolar diffusion. Ohmic resistivity dominates in dense
regions such as the mid-plane of the inner disk, while ambipolar diffusion dominates in tenuous
gas near the disk surface and in the outer disk \citep[for a review, see, e.g.,][]{armitage11}. The ambipolar
diffusion dominated outer disk is the
focus of the current study. In the ambipolar regime, ions and electrons are tied to magnetic fields and are
collisionally coupled to the predominantly neutral gas.  If the collisions are infrequent enough, this
effect acts to damp out MRI turbulence. In almost all models, the outer disk has most of the mass and the
longest viscous time scale, and hence understanding the dynamics of this region is critical for studies of
disk evolution.

In a previous paper, \cite{simon13a} (Paper I), we studied how ambipolar diffusion
affected MRI turbulence in vertically stratified disk models, {\em assuming} that there was no net flux of
vertical magnetic field threading the disk gas. We first carried out a series of local,
shearing box simulations with a temporally and
spatially constant collision to dynamical frequency ratio (referred to here as the ambipolar Elsasser
number). In agreement with unstratified simulations \citep{bai11a},
we found that as the Elsasser number decreases (i.e., ambipolar diffusion becomes stronger), MRI turbulence
weakens, but not with the same monotonic behavior as was observed in the vertically {\it unstratified}
simulations of \cite{bai11a}.

We then simulated the MRI in realistic disk models, with a vertical profile of ambipolar diffusion based
upon the far ultraviolet (FUV) ionization model of \cite{perez-becker11b}. This
ionization model assumes that FUV photons penetrate a thin surface layer of the disk (down to a column
of $\sim $0.01--0.1 g cm$^{-2}$), and completely ionize trace species such as carbon and sulfur.
Below the FUV ionization layer, chemical disk model calculations \citep{bai11b,bai11c} suggest an
ambipolar Elsassar number that is of the order of unity (or lower), which damps the MRI. With this
profile, our zero net flux simulations developed a layered structure analogous to the Ohmic dead
zone model of \cite{gammie96}. We found vigorous surface accretion within the FUV ionization layer,
overlying an ``ambipolar damping zone" near the mid-plane, where the MRI was inactive.

Observations of protoplanetary disk lifetimes and T~Tauri accretion rates \citep{hartmann98a} are broadly
consistent with a \cite{shakura73} $\alpha \sim 10^{-2}$, which happens to be close to
the value derived from {\em ideal} MHD simulations of the MRI in the zero net vertical flux limit \citep{davis10,simon12}. This implies that any
significant suppression of the MRI, due to non-ideal effects in the outer disk, is liable to conflict with
observations, and indeed we found that the ambipolar damping zone seen in our zero net flux simulations
failed to yield accretion rates as large as those typically observed (by an order of magnitude or more). We
thus suggested that a net flux might be a prerequisite for generating higher accretion rates.

In this companion paper, we study the effect of a non-zero net vertical field on MRI-driven
turbulence in the outer regions of protoplanetary disks. Will the presence of a net vertical magnetic flux
enhance turbulence levels enough to agree with observational constraints?  Will this same net flux contribute
to any vertical angular momentum transport through a magnetic wind, such as that seen in recent studies \cite[e.g.,][]{fromang13}?  We explore several field
strengths of the net field and quantify both the disk structure and turbulence levels (from which we ultimately
estimate accretion rates) in order to answer these questions.

The structure of the paper is as follows.  In Section~\ref{method}, we describe our equations and numerical
algorithm, the ionizational model that we employ, and the initial conditions for our simulations.  We also
carry out a convergence study of vertically stratified, net vertical field MRI simulations in order to
justify the resolution that we employ in the rest of the paper.   We then consider the nature of the MRI
under the influence of the layered ionization structure in Section~\ref{results}. Section~\ref{discussion}
discusses the implications of our results for real protoplanetary disks, and we wrap up with conclusions in
Section~\ref{conclusions}.

\newpage
\section{Method}
\label{method}

We have carried out a series of simulations in the local, shearing box approximation \citep{hawley95a}, with an isothermal equation of state, and with outflow vertical boundary conditions that have been modified to enhance the buoyant removal of magnetic flux from the domain.  These shearing box simulations are located at large radial distances from the central star and contain a highly simplified ionization model in which a thin layer above and below the disk mid-plane is assumed to be very strongly ionized due to stellar FUV photons \citep{perez-becker11b}; below these highly ionized layers, we assume a constant, yet large value for the strength of ambipolar diffusion.  We now describe our calculations in more detail.

\subsection{Numerical Method}
\label{num_method}

As in Paper I, we use \textit{Athena}, a second-order accurate Godunov
flux-conservative code for solving the equations of MHD. 
\textit{Athena} uses the dimensionally unsplit corner transport upwind (CTU) method
of \cite{colella90} coupled with the third-order in space piecewise
parabolic method (PPM) of \cite{colella84} and a constrained transport
\citep[CT;][]{evans88} algorithm for preserving the $\del \cdot {\bmath
B}$~=~0 constraint.  We use the HLLD Riemann solver to calculate the
numerical fluxes \cite[]{miyoshi05,mignone07b}.  A detailed description
of the base \textit{Athena} algorithm and the results of various test problems
are given in \cite{gardiner05a}, \cite{gardiner08}, and \cite{stone08}.

We again take advantage of the shearing box approximation in order to better resolve
small scales where ambipolar diffusion becomes important.  The shearing box is a model
for a local, co-rotating disk patch whose size is small compared to the
radial distance from the central object, $R_0$.  This allows the
construction of a local Cartesian frame $(x,y,z)$ that is defined in terms of the disk's
cylindrical co-ordinates $(R,\phi,z^\prime)$ via  $x=(R-R_0)$, $y=R_0 \phi$, and $z = z^\prime$.
The local patch  co-rotates with an angular velocity $\Omega$ corresponding to
the orbital frequency at $R_0$, the center of the box; see \cite{hawley95a}.  The equations to solve are:

\begin{equation}
\label{continuity_eqn}
\frac{\partial \rho}{\partial t} + \del \cdot (\rho {\bmath v}) = 0,
\end{equation}
\begin{equation}
\label{momentum_eqn}
\begin{split}
\frac{\partial \rho {\bm v}}{\partial t} + \del \cdot \left(\rho {\bm v}{\bm v} - {\bm B}{\bm B}\right) + \del \left(P + \frac{1}{2} B^2\right) \\
= 2 q \rho \Omega^2 {\bm x} - \rho \Omega^2 {\bm z} - 2 {\bm \Omega} \times \rho {\bm v} \\
\end{split}
\end{equation}
\begin{equation}
\label{induction_eqn}
\frac{\partial {\bmath B}}{\partial t} - \del \times \left({\bmath v} \times {\bmath B}\right) = \del \times \left[\frac{({\bmath J}\times {\bmath B})\times {\bmath B}}{\gamma \rho_i \rho}\right],
\end{equation} 

\noindent 
where $\rho$ is the mass density, $\rho {\bmath v}$ is the momentum
density, ${\bmath B}$ is the magnetic field, $P$ is the gas pressure,
and $q$ is the shear parameter, defined as $q = -d$ln$\Omega/d$ln$R$.
We use $q = 3/2$, appropriate for a Keplerian disk.  For simplicity and numerical convenience, we
assume an isothermal equation of state $P = \rho \cs^2$, where $\cs$
is the isothermal sound speed.  From left to right, the source terms
in equation~(\ref{momentum_eqn}) correspond to radial tidal forces
(gravity and centrifugal), vertical gravity, and the Coriolis force. The source
term in equation~(\ref{induction_eqn}) is the effect of ambipolar diffusion
on the magnetic field evolution, where $\rho_i$ is the ion density, and
$\gamma$ is the coefficient of momentum transfer for ion-neutral
collisions. Note that our system of units has the magnetic permeability $\mu = 1$, and
the current density is

\begin{equation}
\label{current}
{\bmath J} = \del \times {\bmath B}.
\end{equation}

Numerical algorithms for integrating these equations are described in detail in
\cite{stone10} (see also the Appendix of \citealp{simon11a}). 
The $y$ boundary conditions are strictly periodic, whereas the $x$ boundaries
are shearing periodic \cite[]{hawley95a}. The electromotive forces (EMFs) at
the radial boundaries are properly remapped to guarantee that the net
vertical magnetic flux is strictly conserved to machine precision using CT
\citep{stone10}. 

As in Paper I, ambipolar diffusion is implemented in a first-order operator-split
manner using CT to preserve the divergence free condition
with an additional step of remapping $J_y$ at radial shearing-box boundaries.
The super time-stepping (STS) technique of \cite{alexiades96} has been implemented to
accelerate our calculations (see the Appendix of Paper I).

The only algorithmic difference between these simulations and those in Paper I lie with the vertical boundaries. The simulations in Paper I used the
outflow boundaries described in \cite{simon11a}.  We observed significant
buildup of magnetic field near the vertical boundaries, which was a result of the density floor employed
(described below) being too large; the magnetic field was not able to buoyantly rise out of the disk.  In this work,
we have modified the vertical boundary conditions for the horizontal magnetic field components in order to circumvent this problem.
In particular, for the upper vertical boundary, the value of $B_i$ (where $i = x,y$) in grid cell $k$ is

\begin{equation}
\label{upper_b_bc}
B_i(k) = B_i(ke) {\rm exp}\left(-\frac{z(k)^2-z(ke)^2}{H^2}\right),
\end{equation}

\noindent
where $ke$ refers to the last physical zone at the upper boundary.  Thus, $B_i(ke)$ is the value of $B_i$ at $ke$ at any given time step.
 An equivalent expression holds for the lower vertical boundary. $B_z$ is extrapolated into the ghost zones using the value in the last physical zone with a standard zero gradient outflow boundary.  We have found that this modification induces the removal of magnetic flux away from the mid-plane and out of the domain.

\subsection{Am Profiles}
\label{am_profiles}

The strength of ambipolar diffusion is characterized by the
ambipolar Elsasser number
\begin{equation}
\label{am1}
{\rm Am}\equiv\frac{\gamma\rho_i}{\Omega},
\end{equation}

\noindent
which corresponds to the number of times a neutral molecule collides with the ions in a
dynamical time ($\Omega^{-1}$).

As in Paper I, we adopt the minimum-mass solar nebular (MMSN) disk model with
$\Sigma=1700R_{\rm AU}^{-3/2}$g cm$^{-2}$ \citep{weidenschilling77,hayashi81}, where
$R_{\rm AU}$ is the disk radius measured in AU. We choose the Am profile based on the far ultraviolet (FUV) ionization model of \citet{perez-becker11b}, where FUV photons strongly ionize a column density of $\Sigma_i \sim0.01-0.1$~g~cm$^{-2}$. The corresponding value of Am within the FUV ionized layer can be expressed as follows \citep{bai13b}

\begin{equation}
\label{Am_FUV}
{\rm Am_{\rm FUV}} \approx3.3\times10^7
\bigg(\frac{f}{10^{-5}}\bigg)\bigg(\frac{\rho}{\rho_{0,{\rm mid}}}\bigg)R_{\rm AU}^{-5/4}\ ,
\end{equation}

\noindent
where $f$ is the ionization fraction and $\rho_{0, {\rm mid}}$ is the mid-plane density.
For simplicity, we fix $f=10^{-5}$. We explore two different
ionization depths $\Sigma_i = 0.01$ g cm$^{-2}$ and $\Sigma_i = 0.1$ g cm$^{-2}$. 
For the larger ionization depth, we conduct simulations that correspond to radial
locations at $R=30$ AU and $R=100$ AU.  For the lower ionization depth, we only consider
$R=30$ AU.

In Paper I, the base of the FUV ionization layer was set to be at a fixed height
assuming that the density profile follows hydrostatic equilibrium (Gaussian). This
assumption was justified in these zero net vertical flux runs due to the
extremely weak level of MRI turbulence. However, with net flux, the density
can differ substantially from Gaussian due to enhanced magnetic pressure
support. Here, we identify the
location of the base of the FUV ionization layer ($z_t$and $z_b$ for top and
bottom, respectively) by integrating at each time step the horizontally averaged mass density
from the boundary towards the mid-plane until $\Sigma_i$ is reached. We then
use Equation (\ref{Am_FUV}) to set the strength of ambipolar diffusion in the
ionized surface layers of the disk. In the mid-plane region $(z_b<z<z_t)$, we simply set Am~=~1. 
This is a good approximation if there
are no very small (PAH-sized) grains in this region \citep{bai11c}. If PAHs
are present, the expected value of Am is dependent on both the grain abundance
and on the magnetic field strength. In this limit our adopted value of Am~=~1
remains a reasonable approximation provided that the {\it turbulent} $\beta \sim 10^2$, a value
that is self-consistent with what we obtain in some of our runs. We note
that it is conceivable that there could be circumstances in which grains
are present and $\beta \gg 10^2$. This would result in ${\rm Am} \ll 1$
and stronger ambipolar damping near the mid-plane. We cannot study this
regime numerically, however, because the Courant-limited diffusive time step
would be too small.

From these considerations, the value of Am changes quite dramatically from Am~=~1
to Am $\sim 10^4$ at the base of the FUV layer. This very large transition
is smoothed over roughly 7 grid zones so as to prevent a discontinuous transition in Am.
The smoothing functions we apply are based upon the error function (ERF).   Thus, the
complete profile of Am for these runs is given by

\begin{equation}
\label{amc}
\small
{\rm Am} \equiv \left\{ \begin{array}{ll}
 {\rm Am_{\rm FUV}} & \quad 
\mbox{$z \ge z_t + \Delta z$} \\
1 + \frac{1}{2}{\rm Am_{\rm FUV}}S^+(z)  & \quad
\mbox{$z_t - n\Delta z < z < z_t + \Delta z$} \\
1 & \quad
 \mbox{$z_b+n\Delta z \le z \le z_t - n \Delta z$} \\
1 + \frac{1}{2}{\rm Am_{\rm FUV}}S^-(z) & \quad
\mbox{$z_b-\Delta z < z < z_b + n\Delta z$} \\
{\rm Am_{\rm FUV}} & \quad
\mbox{$z \le z_b - \Delta z$}
\end{array} \right.
\end{equation}

\noindent
where $S^+(z)$ and $S^-(z)$ are the smoothing functions defined as 
 
\begin{equation}
\small
\label{splus}
S^+(z) \equiv 1+{\rm ERF}\left(\frac{z-0.9z_t}{\Delta z}\right),
\end{equation}
\begin{equation}
\small
\label{sminus}
S^-(z) \equiv 1-{\rm ERF}\left(\frac{z-0.9z_b}{\Delta z}\right),
\end{equation}

\noindent
Here, $n = 8$ and $\Delta z = 0.05 H$.  These numbers were chosen to give a reasonably
resolved transition region between Am = 1 and ${\rm Am_{\rm FUV}}$. We also note that
since $Am_{\rm FUV}\gg1$ in the above formula, the FUV photons effectively penetrate slightly
deeper than $z_t$ and $z_b$ by about $0.2H$.

\subsection{Simulations}

We have run a series of simulations both in the ideal MHD limit (used for convergence studies;
see below), and using the Am profiles described in the previous section. Aside from the Am profile,
all simulations start from the same initial conditions.  The gas density is set to be in hydrostatic
equilibrium for an isothermal gas,

\begin{equation}
\label{density_init}
\rho(x,y,z) = \rho_0 {\rm exp}\left(-\frac{z^2}{H^2}\right),
\end{equation}
where $\rho_0 = 1$ is the mid-plane density, and $H$ is the scale height in the disk,

\begin{equation}
\label{scale_height}
H = \frac{\sqrt{2} \cs}{\Omega}.
\end{equation}
The isothermal sound speed, $\cs = 7.07 \times 10^{-4}$, corresponding to an initial value for
the mid-plane gas pressure of $P_0 = 5\times 10^{-7}$.  With $\Omega = 0.001$, the value for
the scale height is $H = 1$.   A density floor of $10^{-4}$ is applied to the physical domain as
too small a density leads to a large $\alf$ speed and a very small
time step.  Furthermore, numerical errors make it difficult
to evolve regions of very small plasma $\beta$ (ratio of thermal pressure to
magnetic pressure).  

\begin{deluxetable*}{l|ccccccccc}
\tabletypesize{\small}
\tablewidth{0pc}
\tablecaption{Shearing Box Simulations\label{tbl:sims}}
\tablehead{
\colhead{Label}&
\colhead{Ambipolar Diffusion}&
\colhead{$\beta_0$}&
\colhead{$z_{\rm bw}$\tablenotemark{$\ast$}}&
\colhead{$\alpha_{\rm total}$\tablenotemark{$\ast$}}&
\colhead{$\alpha_{\rm turb}$\tablenotemark{$\dagger$}}&
\colhead{$\alpha_{\rm bw}$\tablenotemark{$\ast$}}&
\colhead{$2\overline{|W_{z\phi}|}_{\rm total}$\tablenotemark{$\ast$}}&
\colhead{$2\overline{|W_{z\phi}|}_{\rm bw}$\tablenotemark{$\ast$}} \\
\colhead{ }&
\colhead{($\Sigma_{\rm i}$ is ionization column)}&
\colhead{ }&
\colhead{$(H)$}&
\colhead{ }&
\colhead{ }&
\colhead{ }&
\colhead{ }&
\colhead{ }&
\colhead{ } } 
\startdata
36Num & none & $10^4$& -- & 0.083 & -- & -- & -- & --  \\
72Num & none & $10^4$& -- & 0.10 & -- & -- & -- & --  \\
144Num & none & $10^4$ & -- & 0.12 & -- & -- & -- & --   \\
AD30AU1e3 & 30 AU, $\Sigma_{\rm i} = 0.1$ g cm$^{-2}$ & $10^3$& $1.83$ & 0.077 & 0.0096 & 0.025 & 0.017 & 0.029 \\
AD30AU1e4 & 30 AU, $\Sigma_{\rm i} = 0.1$ g cm$^{-2}$ & $10^4$ & $2.17$ & 0.020 & 0.011 & 0.013 & 0.0015 & 0.0038 \\
AD30AU1e5 & 30 AU, $\Sigma_{\rm i} = 0.1$ g cm$^{-2}$ & $10^5$ & $2.40$ &  0.0033 & 0.0022 & 0.0022 & 0.00019 & 0.00055\\
AD30AU1e4L& 30 AU, $\Sigma_{\rm i} = 0.01$ g cm$^{-2}$ & $10^4$ & $2.26$ & 0.0075& 0.0024 & 0.0033 & 0.0020 & 0.0038\\
AD30AU1e5L & 30 AU, $\Sigma_{\rm i} = 0.01$ g cm$^{-2}$ & $10^5$ & $2.47$  & 0.0021& 0.0012 & 0.0012 & 0.00019 & 0.00044\\
AD100AU1e4 & 100 AU, $\Sigma_{\rm i} = 0.1$ g cm$^{-2}$  & $10^4$ & $2.22$ & 0.030 & 0.020 & 0.023 & 0.0014 & 0.0046 \\
AD100AU1e5 & 100 AU, $\Sigma_{\rm i} = 0.1$ g cm$^{-2}$ & $10^5$ & $2.33$  & 0.0052 & 0.0041 & 0.0039 & 0.00014 & 0.00059  \\
\enddata
\tablenotetext{$\ast$}{\scriptsize These quantities are described in Section~\ref{results_wind}.}
\tablenotetext{$\dagger$}{\scriptsize $\alpha_{\rm turb}$ is calculated by subtracting off the large scale Maxwell stress from $\alpha$: $\alpha -\overline{\left(-\langle B_x\rangle\langle B_y\rangle/\langle \rho \cs^2\rangle\right)}$, where the overbar denotes a time average.}
\end{deluxetable*}

In all runs, the initial magnetic field is a net vertical field. It is well known that for purely
vertical field, the MRI sets in from a transient channel flow. For relatively strong net vertical
magnetic flux, the channel flow is so strong as to cause numerical problems and/or disk
disruption in the simulations \citep{miller00}. To circumvent such potential difficulties, we
add a sinusoidal varying vertical field on top of a purely vertical field

\begin{equation}
\label{initi_b}
B_z = B_0 \left[1+\frac{1}{2}{\rm sin}\left(\frac{2\pi}{L_x}x\right)\right],
\end{equation}
where $L_x$ is the domain size in the $x$ dimension, and 

\begin{equation}
\label{bo}
B_0 = \sqrt{\frac{2P_0}{\beta_0}}.
\end{equation}
Here, $B_0$ is the net vertical magnetic field, characterized by $\beta_0$, the ratio of gas
pressure to the magnetic pressure of the net vertical field at the disk mid-plane.  It is a free
parameter in our simulations. With the asymmetry introduced by the extra sinusoidal variation in the vertical field, the strong growth
of channel flows \cite[see][]{hawley95a,miller00} is suppressed at early stages, and the simulation can integrate beyond the
initial transient without numerical problems. All other magnetic field components are initialized to be zero.

To seed the MRI, random perturbations are added to the density and velocity components.  The amplitude of these perturbations are $\delta\rho/\rho_0$= 0.01 and
$\delta v_i = 0.004 \cs$ for $i=x,y,z$. 

All of our simulations are listed in Table~\ref{tbl:sims}, along with some properties of each simulation.
We have carried out three ideal MHD (i.e., no ambipolar diffusion) simulations, each at a different resolution.  These runs are labelled as RNum where R is the number of grid zones per $H$; R = 36, 72, and 144 here.  Each of these runs are initialized with $\beta_0 = 10^4$ and have domain size $L_x = 2H$ in the $x$ direction, $L_y = 4H$ in the $y$ direction, and $L_z = 8H$ in the $z$ direction. The ambipolar calculations are prefixed with AD, followed by the radial location of the shearing box (e.g., 30AU) and the value of $\beta_0$ (e.g., 1e3).  There are also two runs that have L at the end of their run name, denoting that they have a lower ionization depth than the other runs (i.e., $\Sigma_i = 0.01$~g~cm$^{-2}$ instead of $\Sigma_i = 0.1$~g~cm$^{-2}$).

All of the ambipolar diffusion calculations are run with a domain size $L_x = 8H$ in the $x$ direction, $L_y = 16H$ in the $y$ direction, and $L_z = 8H$ in the $z$ direction.  We choose this domain size based upon the results of Paper I, where we found that a large domain size was necessary to accurately capture the physics of the MRI in the presence of ambipolar diffusion.\footnote{In Paper I, we used a box size of $(L_x, L_y, L_z)H = (4,8,8)H$ for most of our runs, but pointed out that for Am $\leq 10$, a larger domain is preferred.}

\begin{figure}
\begin{center}
\includegraphics[width=0.48\textwidth,angle=0]{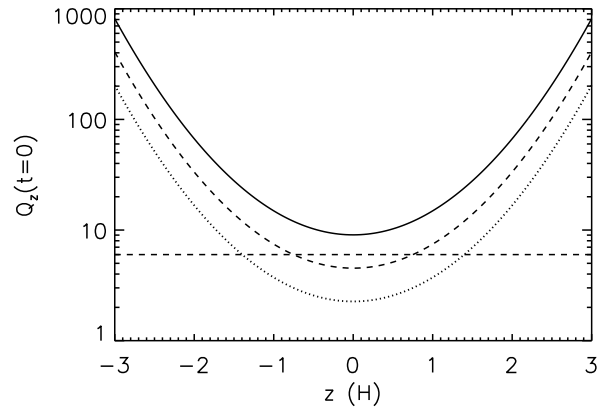}
\end{center}
\caption{
Initial vertical profile of $Q_z$ as defined by equation~(\ref{qz}) for the three ideal MHD runs with
$\beta_0=10^{4}$.  The dotted line corresponds to a resolution of 36 zones per $H$, the dashed line is 72 zones per $H$, and the solid line is 144 zones per $H$.  The horizontal dashed line is $Q_z = 6$, which is the value below which the MRI will not be sufficiently resolved according to the results of \cite{sano04}.  The lowest resolution run has a significant vertical region in which $Q_z < 6$, whereas $Q_z < 6$ in a very small vertical region for the 72 zones per $H$ run.  There is no such region for the highest resolution simulation.
}
\label{qz_z_initial}
\end{figure}
 
\subsection{Convergence}
\label{ideal}

In this section, we describe our ideal MHD calculations designed to study the convergence of turbulent saturation with numerical resolution.  To motivate these
calculations, let us first consider the quality factor in the $z$ direction, $Q_{z}$,  which returns the number of grid zones per characteristic MRI wavelength of the background vertical magnetic field.

\begin{equation}
\label{qz}
Q_z \equiv \frac{\lambda_{{\rm MRI},z}}{\Delta z} = \frac{2\pi \cs}{\Omega\Delta z}\sqrt{\frac{2}{\beta_z}},
\end{equation}

\noindent
where $\Delta z$ is the grid spacing in $z$ and $\beta_z$ is the plasma $\beta$ of the background vertical field as a function of height.
At a resolution of 36 grid zones per $H$ and the three values of $\beta_0$ explored in this work, $\beta_0 = 10^3, 10^4$, and $10^5$, we find that $Q_z = 7.2, 2.3$, and $0.72$ respectively, at $z = 0$ and $t = 0$.  Figure~\ref{qz_z_initial} plots the initial distribution of $Q_z$ for the inner 6$H$ of the ideal MHD simulations.  The horizontal dashed line is $Q_z = 6$, below which the growth of the vertical field MRI is considered to be under-resolved \citep{sano04}.\footnote{It is worth noting that this number is an estimate and should not be thought of as a definite demarcation for where the MRI becomes under-resolved.} Thus, for the two weaker fields ($\beta_0 = 10^4$ and $\beta_0 = 10^5$), the MRI is likely to be under-resolved near the mid-plane.  For $\beta_0 = 10^4$ and at the mid-plane, $Q_z =  4.5$ for 72 zones per $H$, and $Q_z =  9.0$ for 144 zones per $H$.  From these numbers and Fig.~\ref{qz_z_initial}, the characteristic vertical MRI modes near the mid-plane go from being under-resolved to reasonably well-resolved as resolution is increased.

\begin{figure}
\begin{center}
\includegraphics[width=0.49\textwidth,angle=0]{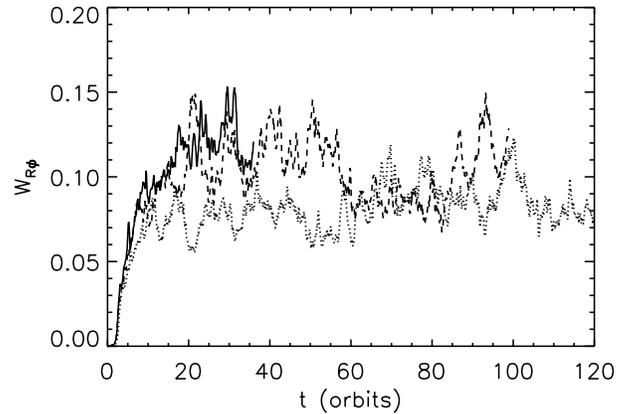}
\end{center}
\caption{
Density-weighted volume average of the total (Maxwell and Reynolds) dimensionless stress versus time for the
ideal MHD simulations.  The dotted line corresponds to the run with 36 zones per $H$, the dashed line is 72 zones per $H$, and the solid line is 144 zones per $H$.
The stress evolutions are very similar, eventually reaching statistically similar values, suggesting that 36 zones per $H$ is a sufficient resolution to capture the basic properties of MRI-driven turbulence in the presence of a weak net vertical field.
}
\label{ideal_res}
\end{figure}

The question we seek to answer is whether or not this critical change in the mid-plane $Q_z$ influences the non-linear outcome of the MRI. The result is shown in Fig.~\ref{ideal_res}, in which we plot the dimensionless, density weighted $R\phi$ component of the stress tensor,

\begin{equation}
\label{wrp}
W_{R\phi} = \frac{\left\langle \rho v_x \delta v_y
- B_xB_y\right\rangle}{\left\langle \rho\cs^2\right\rangle},
\end{equation}

\noindent
where $\delta v_y$ is the $y$ velocity with the orbital shear subtracted.  The angled brackets denote a volume average (this applies here and throughout the rest of the paper).
 The dotted line is the ideal MHD run at 36 zones per $H$, the dashed line is 72/$H$, and the solid line is 144/$H$.  All three calculations reach roughly the same saturation amplitude.

\begin{figure*}
\begin{minipage}[!ht]{8cm}
\begin{center}
\includegraphics[width=1\textwidth,angle=0]{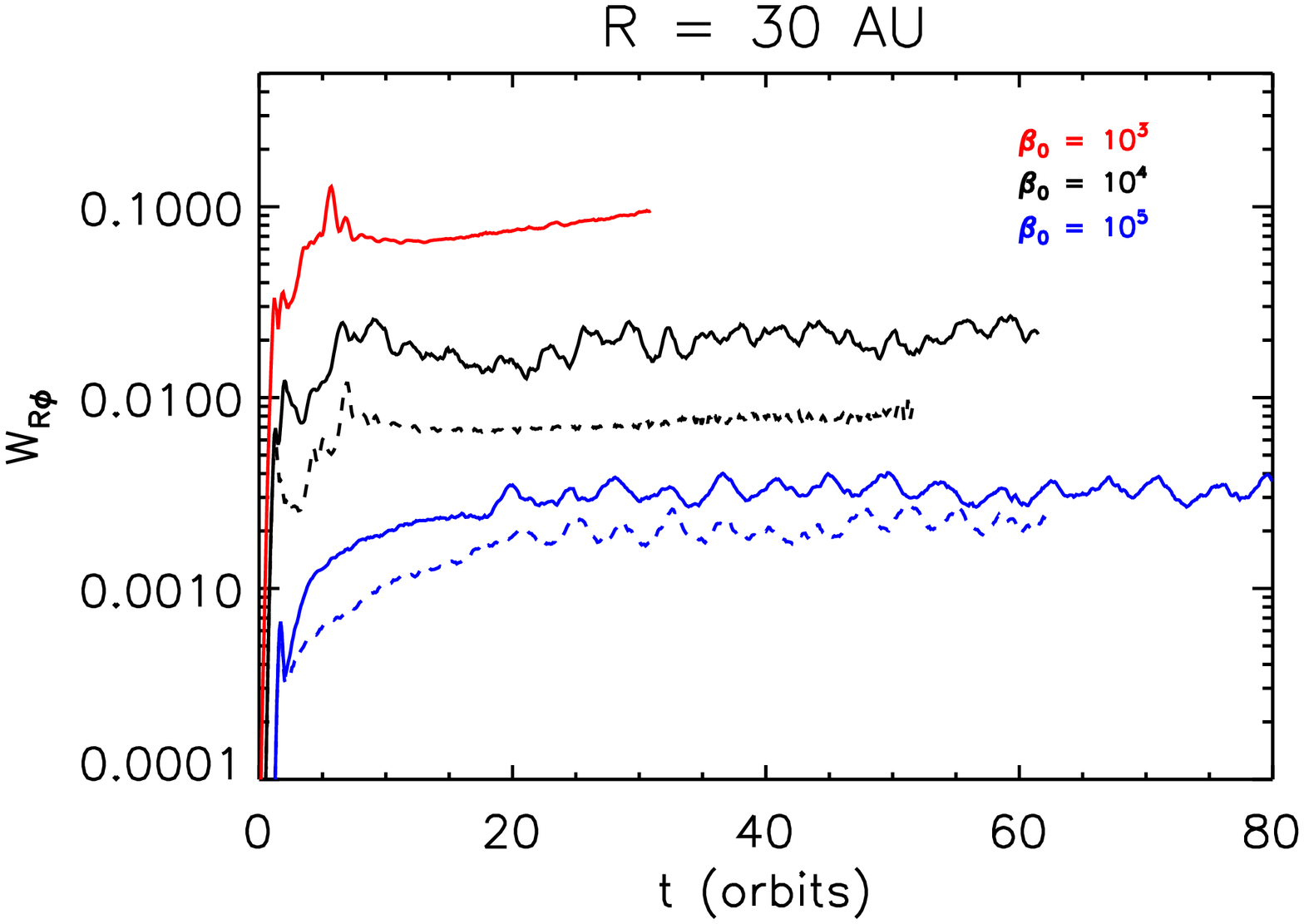}
\end{center}
\end{minipage}
\begin{minipage}[!ht]{8cm}
\begin{center}
\includegraphics[width=1\textwidth,angle=0]{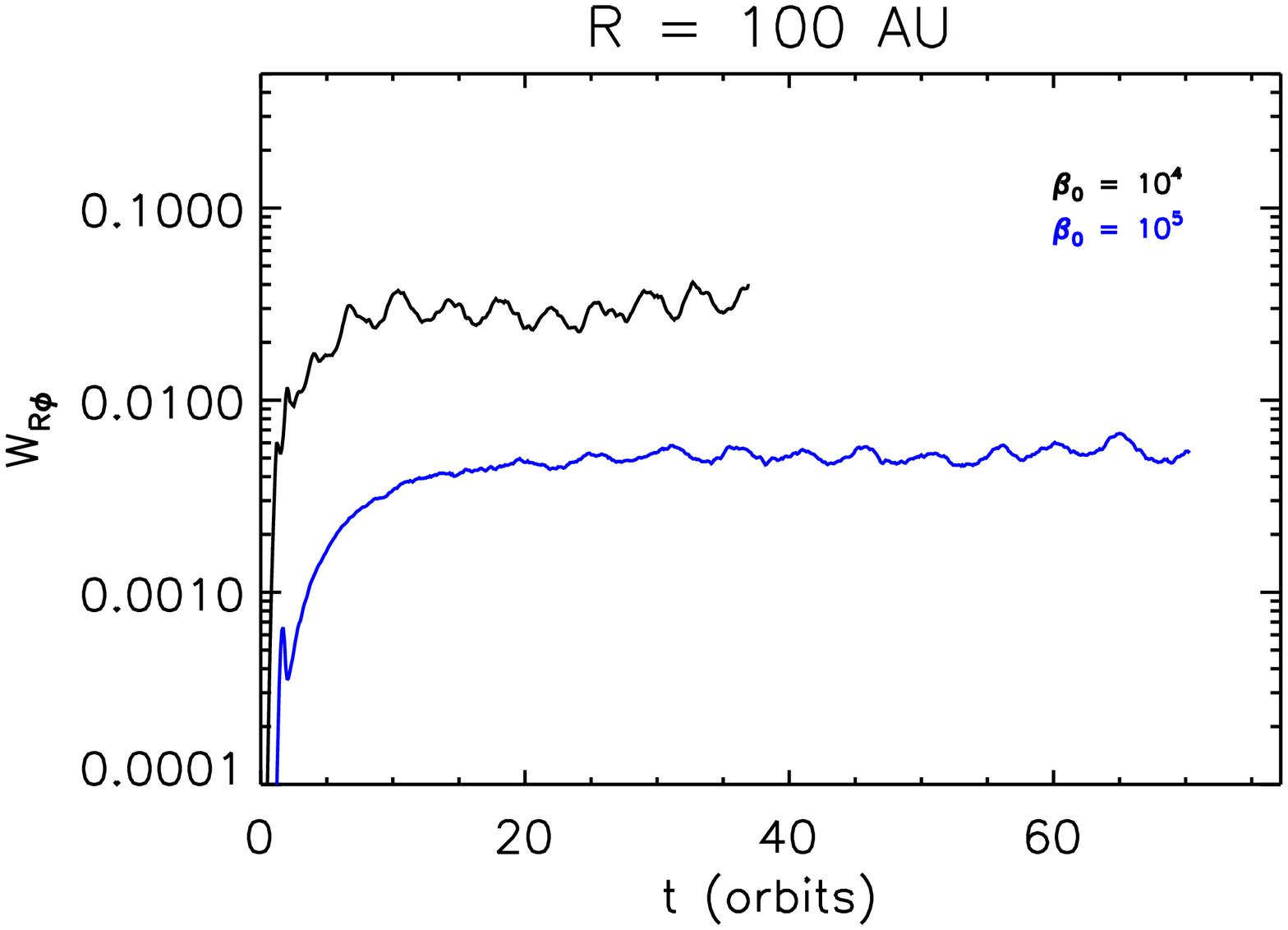}
\end{center}
\end{minipage}
\caption{
Density-weighted volume average of the total (Maxwell and Reynolds) dimensionless stress versus time for the non-ideal MHD simulations with a layered Am profile.  The left panel corresponds to the runs at 30 AU and the right panel corresponds to 100 AU.  As described by the legend, the color of the line represents the strength of the net vertical magnetic field; the red line corresponds to $\beta_0 = 10^3$, the black lines correspond to $\beta_0 = 10^4$, and the blue lines correspond to $\beta_0 = 10^5$.  The solid lines correspond to an ionization depth of $\Sigma_{\rm i} = 0.1$ g cm$^{-2}$, and the dashed lines correspond to $\Sigma_{\rm i} = 0.01$ g cm$^{-2}$.   The stress values all fall within the range 0.001-0.1, and there is a strong dependence of stress on the net vertical magnetic flux.
}
\label{stress}
\end{figure*}

We can time average the normalized stress, thus defining the Shakura-Sunyaev $\alpha$ parameter,

\begin{equation}
\label{alpha}
\alpha \equiv \overline{W_{R\phi}}
\end{equation} 

\noindent where the over-bar denotes a time average (this applies here and throughout the rest of the paper), which is done from orbit 15 onwards for each of the ideal MHD simulations.  The value of $\alpha$ is listed in Table~\ref{tbl:sims}; the time-averaged stresses are all roughly equal, though there is an approximately 20\% increase in $\alpha$ with each doubling of resolution.

\begin{figure*}
\begin{minipage}[!ht]{8cm}
\begin{center}
\includegraphics[width=1\textwidth,angle=0]{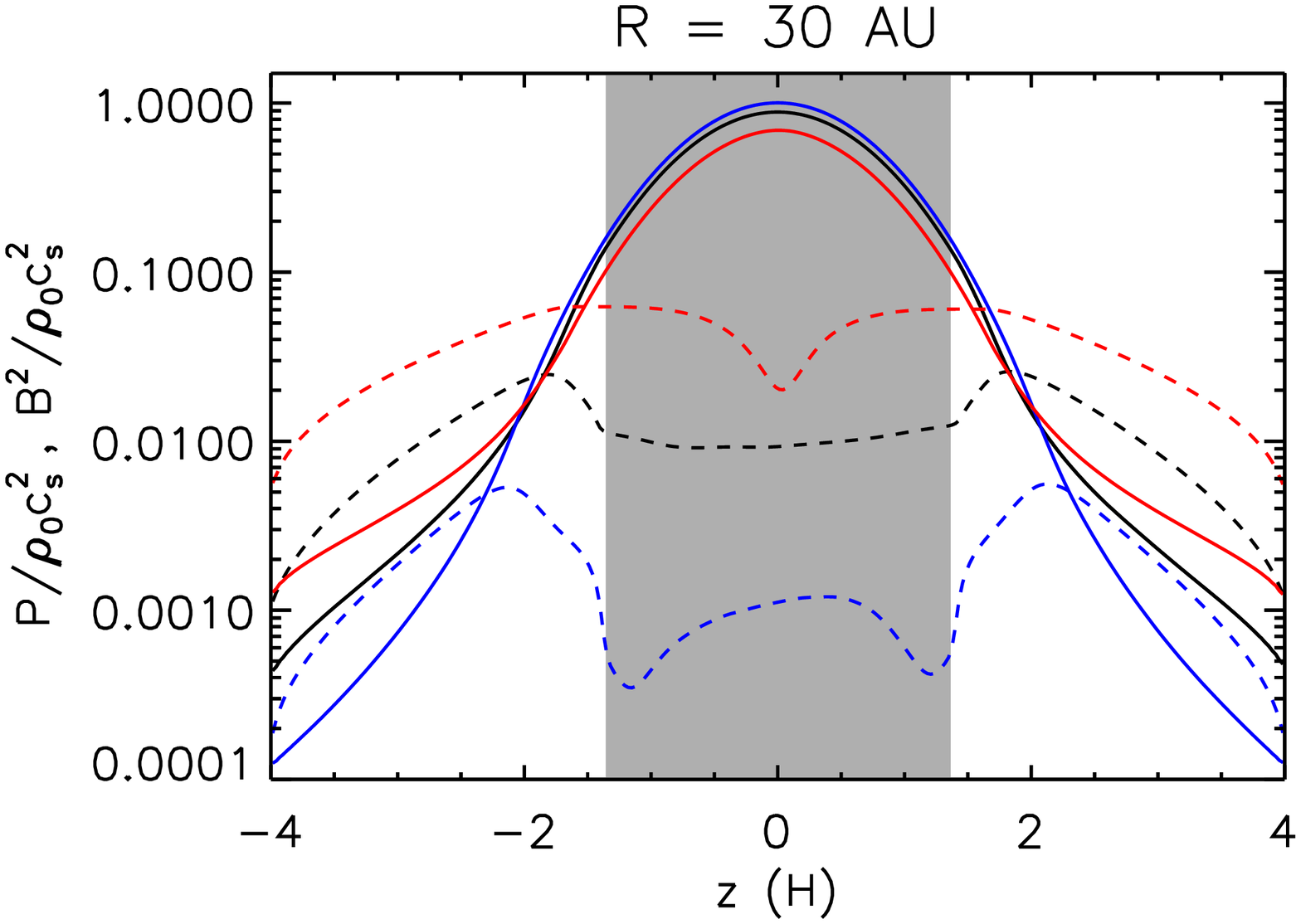}
\end{center}
\end{minipage}
\begin{minipage}[!ht]{8cm}
\begin{center}
\includegraphics[width=1\textwidth,angle=0]{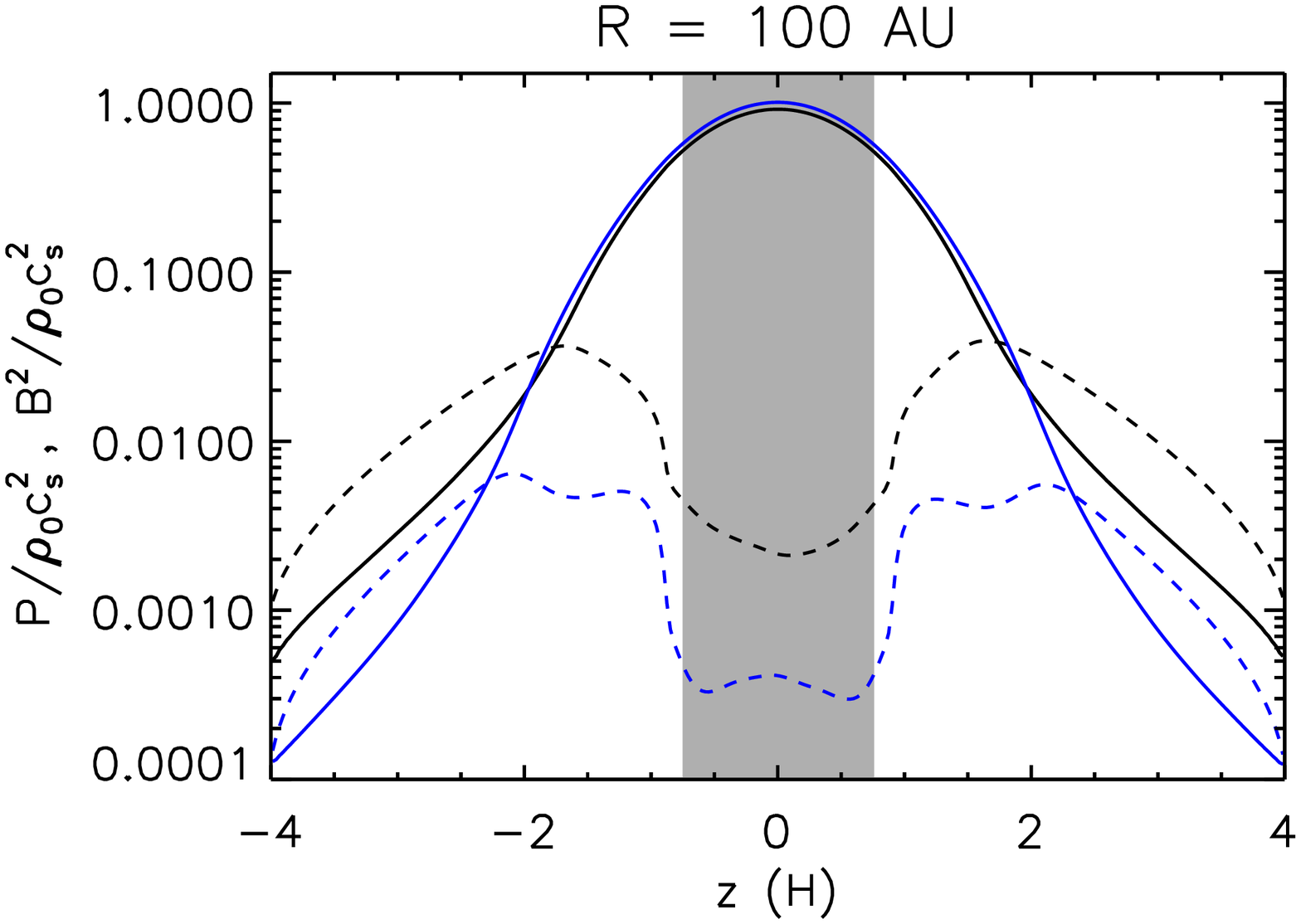}
\end{center}
\end{minipage}
\caption{
Time- and horizontally-averaged vertical profile for gas pressure (solid lines) and magnetic pressure (dashed lines) normalized by the initial mid-plane gas pressure.  The data correspond to the layered Am simulations with $\Sigma_{\rm i} = 0.1$ g cm$^{-2}$, and the left panel corresponds to the runs at 30 AU and the right panel corresponds to 100 AU.  As in Fig.~\ref{stress}, the color of the line represents the strength of the net vertical magnetic field; the red line corresponds to $\beta_0 = 10^3$, the black lines correspond to $\beta_0 = 10^4$, and the blue lines correspond to $\beta_0 = 10^5$. The shaded region denotes where Am = 1. In all cases, the gas pressure maintains a roughly Gaussian profile near the disk mid-plane, but then deviates from Gaussian at large $|z|$ due to magnetic pressure support.  The magnetic energy shows a dip towards the mid-plane.
}
\label{energy}
\end{figure*}

This relatively small difference between averaged stress levels at different resolutions is somewhat surprising because it seems natural for the background vertical field to continually play a role in driving the MRI during the non-linear state.  Indeed, one might expect a more significant change in $\alpha$ as the resolution is doubled. One explanation for this behavior is that as the vertical magnetic field energy is increased (though, the net vertical field remains the same), $Q_z$ (now defined by the turbulent magnetic field, and {\it not} the net vertical field) reaches a regime where the MRI is sufficiently well-resolved at all $z$. Another, more likely, explanation is that the regions where $Q_z \lesssim 6$ are small compared to the rest of the domain.  Therefore, when adding up all the stress in the domain, there isn't much difference between the volume-averaged stresses at different resolutions.  

Finally, all of our non-ideal MHD calculations have Am = 1 near the disk mid-plane. The corresponding most unstable wavelength is nearly twice that of ideal MHD due to the dissipative nature of ambipolar diffusion \citep{Wardle99,KunzBalbus04,bai11a}. 
Altogether, these considerations give us confidence that 36 zones per $H$ is sufficient resolution to model the outer regions of protoplanetary disks.

\begin{figure*}
\begin{minipage}[!ht]{8cm}
\begin{center}
\includegraphics[width=1\textwidth,angle=0]{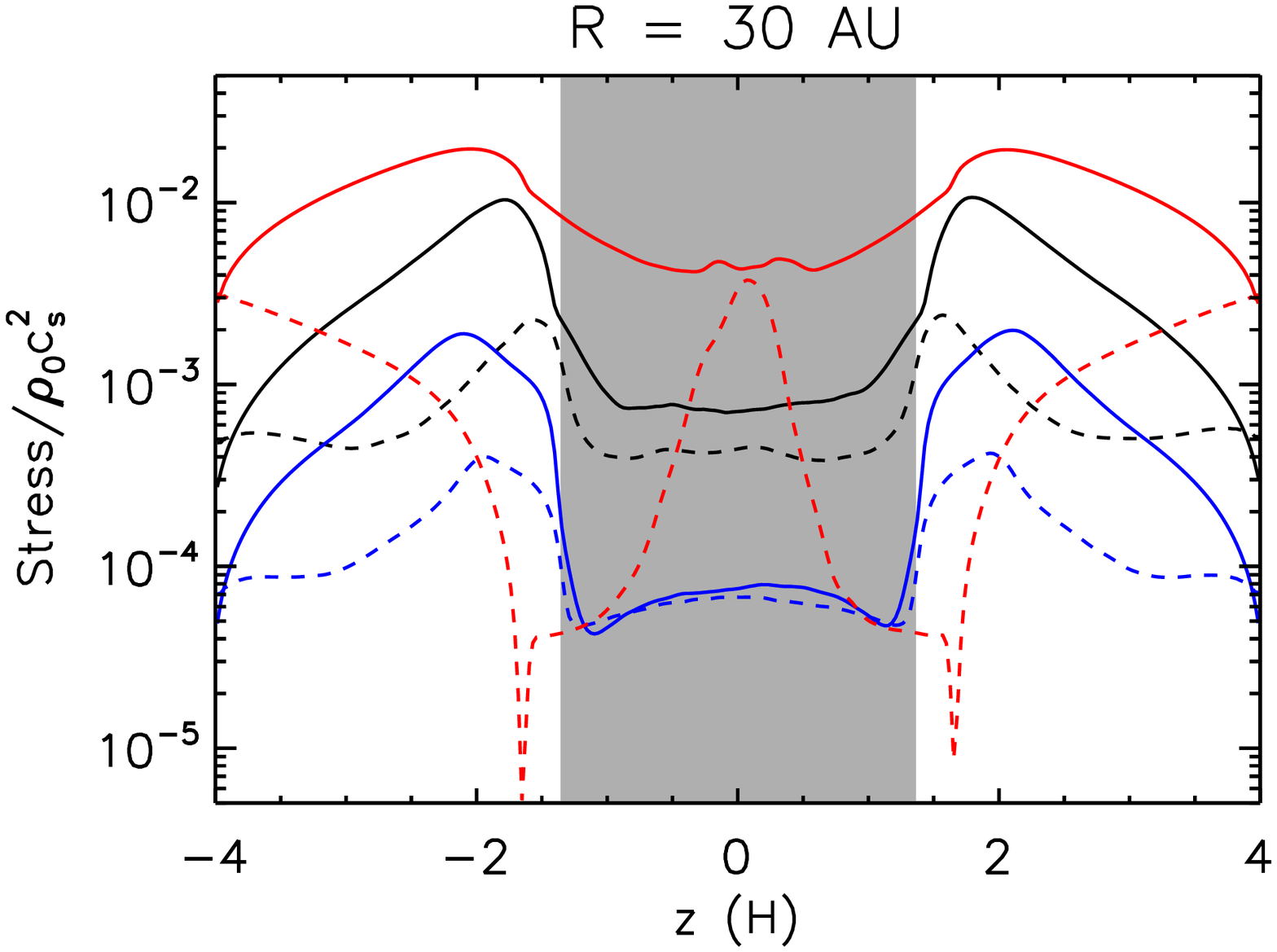}
\end{center}
\end{minipage}
\begin{minipage}[!ht]{8cm}
\begin{center}
\hspace{-3.8cm}
\includegraphics[width=1\textwidth,angle=0]{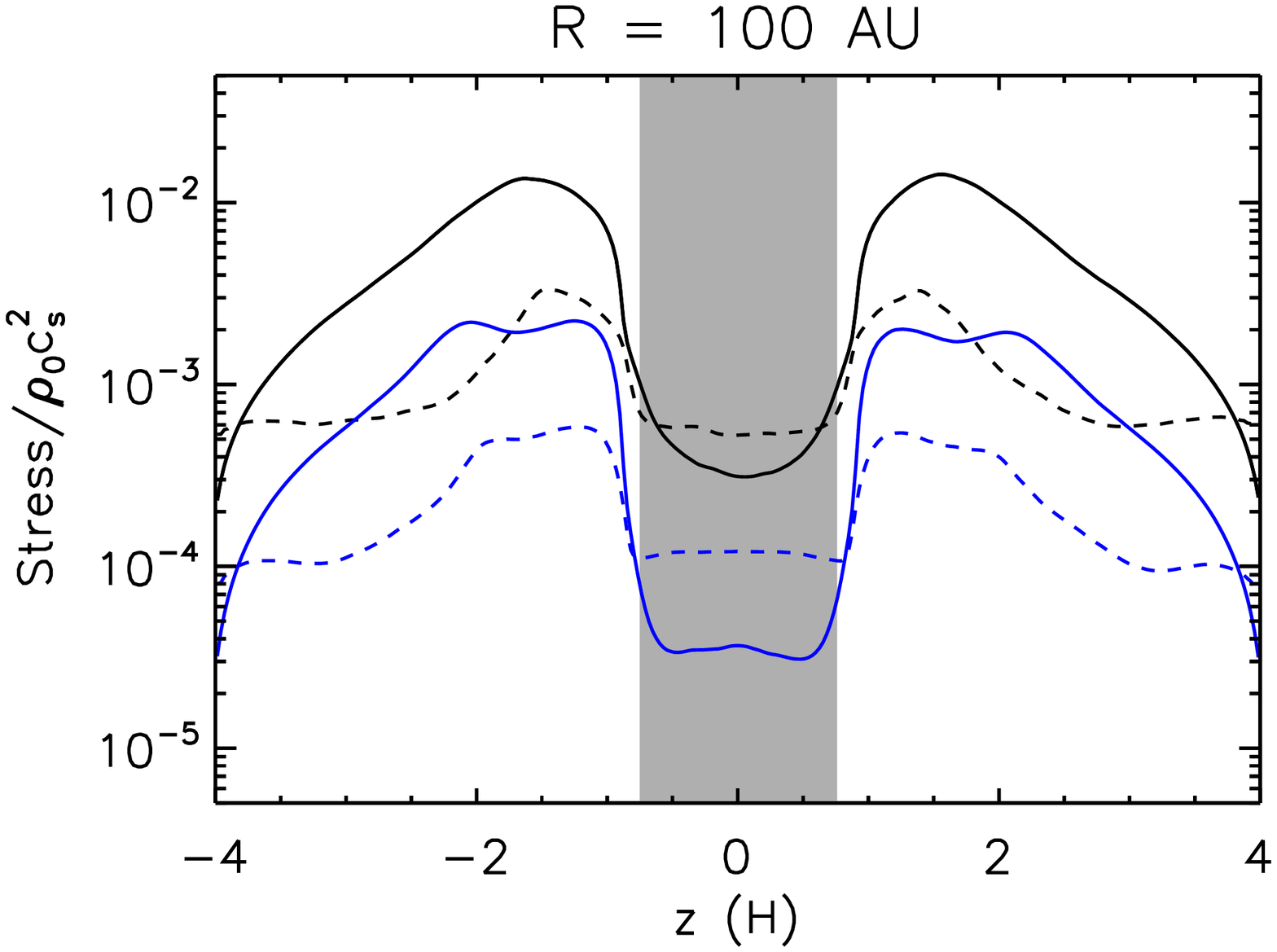}
\end{center}
\end{minipage}
\begin{minipage}[!ht]{8cm}
\begin{center}
\includegraphics[width=1\textwidth,angle=0]{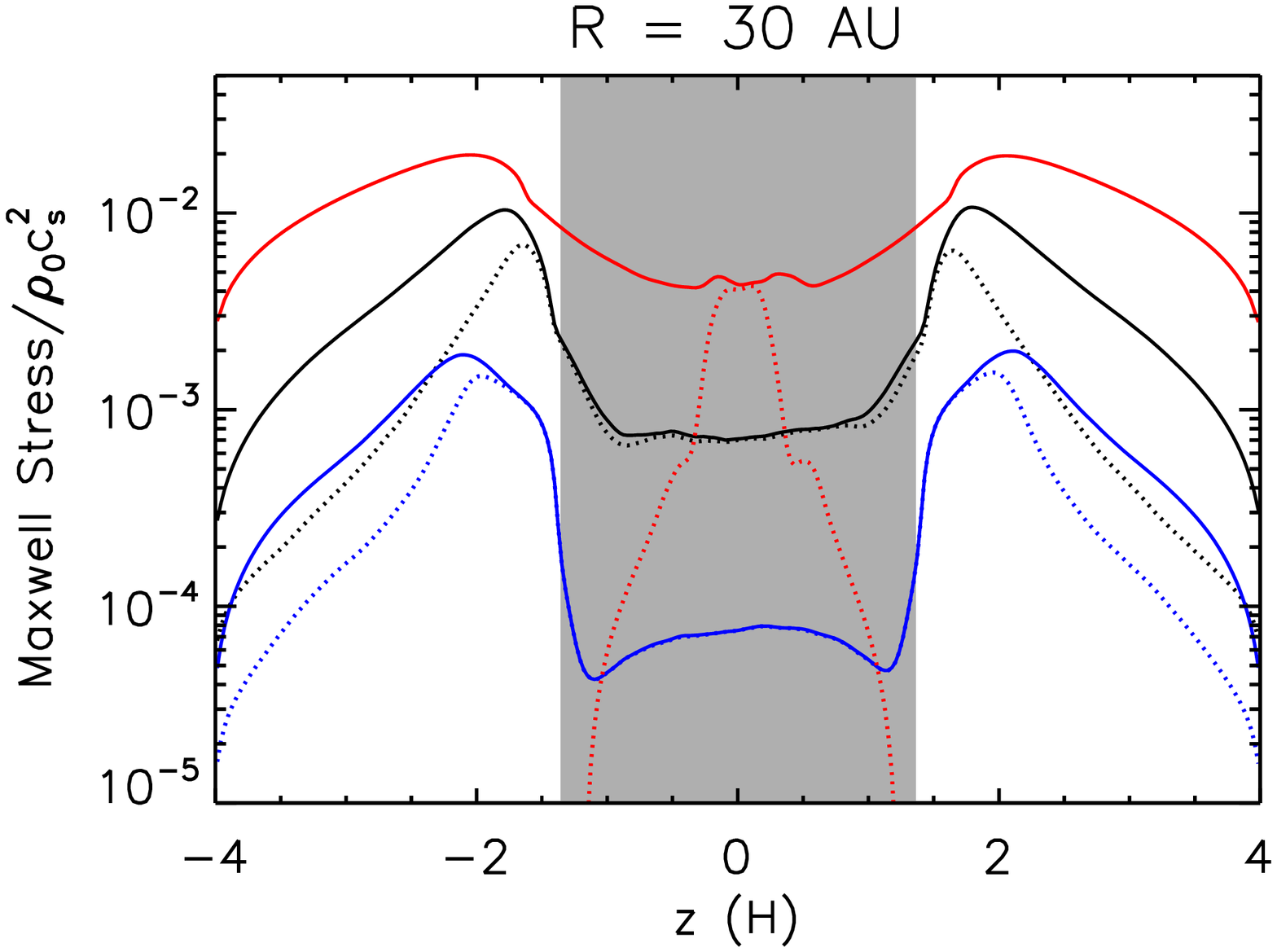}
\end{center}
\end{minipage}
\begin{minipage}[!ht]{8cm}
\begin{center}
\includegraphics[width=1\textwidth,angle=0]{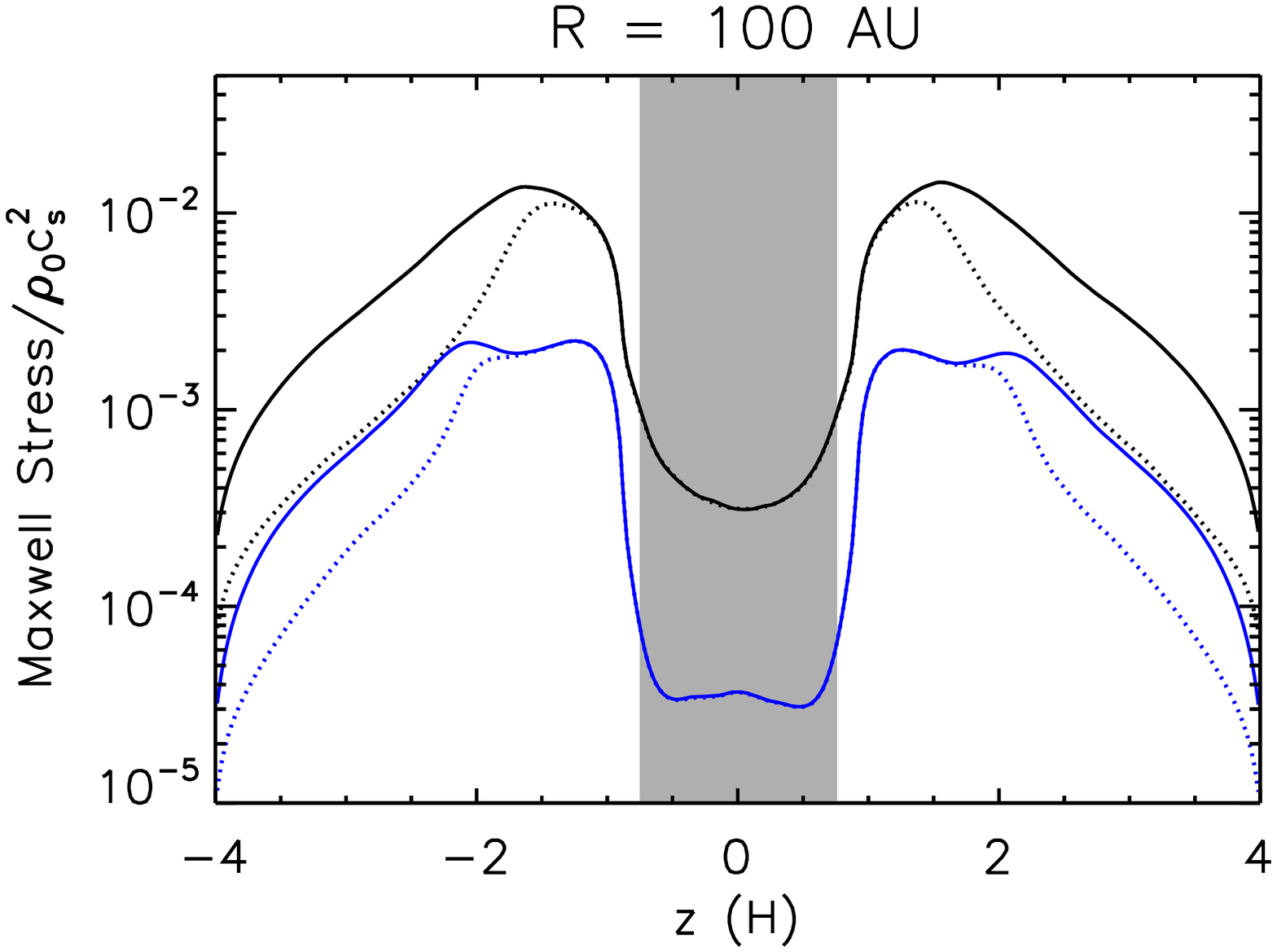}
\end{center}
\end{minipage}
\caption{
Top row: Time- and horizontally-averaged vertical profile for Maxwell stress (solid lines) and Reynolds stress (dashed lines) normalized by the initial mid-plane gas pressure.  Bottom row: Time- and horizontally-averaged vertical profile for total Maxwell stress (solid lines) and the turbulent Maxwell stress (dashed line) calculated by subtracting the largest scale stress $-\langle B_x\rangle\langle B_y\rangle$ from the total Maxwell stress. In all panels, the data correspond to the layered Am simulations with $\Sigma_{\rm i} = 0.1$ g cm$^{-2}$. The left panel corresponds to the runs at 30 AU, and the right panel corresponds to 100 AU.  As in Fig.~\ref{stress}, the color of the line represents the strength of the net vertical magnetic field; the red line corresponds to $\beta_0 = 10^3$, the black lines correspond to $\beta_0 = 10^4$, and the blue lines correspond to $\beta_0 = 10^5$.  The shaded region denotes where Am = 1.  The Maxwell stress dominates over the Reynolds stress in almost all cases; the exception is within the Am = 1 region at 100 AU.  There is a spike in Reynolds stress at $z = 0$ for AD30AU1e3.  The Maxwell stress has a non-negligible large scale component, except for at $z = 0$ for AD30AU1e3 and within $\sim 2H$ of the mid-plane for all of the other runs.
}
\label{stress_z}
\end{figure*}

\section{Results}
\label{results}

We now present the calculations with ambipolar diffusion.  In the first subsection, we present our data via various diagnostics and then go on in subsequent subsections to explain the observed features in more detail.  Finally, as mentioned above, these simulations are listed in Table~\ref{tbl:sims} and are prefixed with an ``AD". For all diagnostics in this section that require time-averaging, the average is done from orbit 10 onwards for AD30AU1e3 and from orbit 20 onwards for all other calculations.  These times were chosen to roughly coincide with the start of the saturated state in which the stress fluctuates about some statistically constant value.

\subsection{Ambipolar Diffusion Simulations}
\label{results_sub}

Figure~\ref{stress} shows the normalized stress evolution for the different vertical field runs at 30 AU (left panel) and 100 AU (right panel).  There is a clear increase in the stress at early times, due to the linear growth of the MRI, followed by turbulent saturation.  Note that once saturation is reached, there does not appear to be any long timescale ($\gtrsim 5$ orbits) fluctuations present in the stress evolution.  There are, however, short timescale, quasi-periodic fluctuations that have a period of roughly 3-5 orbits.  These fluctuations are present for AD30AU1e4, AD30AU1e5, AD30AU1e5L, AD100AU1e4, and AD100AU1e5.  The remaining calculations, AD30AU1e3 and AD30AU1e4L, show a very flat stress evolution after the initial growth phase.

The short evolution time for AD30AU1e3 is a result of significant mass loss during this time;  by orbit 30, roughly 50\% of the mass in the domain has been lost due to a strong vertical outflow (described below).  We did not feel that it would be an accurate representation of a real disk if we integrated this particular run further.  This mass loss is in itself interesting, and we discuss it further in Section~\ref{discussion}.

There is a very strong dependence of the stress on $\beta_0$, consistent with earlier work \cite[e.g.,][]{hawley95a,pessah07,bai13a}.  From Table~\ref{tbl:sims}, $\alpha$ ranges from $\sim 0.001$ to $\sim 0.1$, depending on the strength of the initial magnetic field. Comparing to the $\alpha$ values in Paper I, we find that the presence of a net vertical field enhances the stress levels significantly, particularly for $\beta_0 \lesssim 10^4$.  For $\beta_0 = 10^5$, the values of $\alpha$ are still larger than those without any vertical magnetic flux by a factor of 2-3. 

The dashed lines on the left panel of Fig.~\ref{stress} correspond to the runs with a lower ionization column: AD30AU1e4L and AD30AU1e5L. Despite having an order of magnitude lower value for $\Sigma_i$, the resulting stress levels differ only by a much smaller factor: 3 in the case of the two runs at $\beta_0=10^4$ and $1.5$ for the case of $\beta_0=10^5$.   These small differences will be explained in Section~\ref{results_lowion}.

We next plot the vertical profiles (averaged in the $x$ and $y$ dimensions and in time) of several quantities.  In each plot, the shaded regions correspond to where Am = 1.  The exact $z$ values corresponding to the top and bottom of the Am = 1 region are slightly different in each simulation.  Thus, in order to simplify the plots and display only one shaded region, we averaged these $z$ values across all of the simulations in a given figure and used the averages to determine the borders of the shaded regions. 

Figure~\ref{energy} shows the energy profile for both the 30 AU and 100 AU runs with the same color scheme as in Fig.~\ref{stress}.  We do not include the lower ionization column runs here.  The solid line is the dimensionless gas pressure, and the dashed line is the dimensionless magnetic energy.   As is usual for vertically stratified MRI calculations \cite[e.g.,][]{miller00,simon11a}, the magnetic field is sub-thermal for $|z| \lesssim 2H$.  Outside of this region, the magnetic field is super-thermal, but continues to drop in magnitude away from the disk mid-plane.  The dip in magnetic energy near the disk mid-plane is a result of the lower ionization levels there (Am = 1).

The vertical stress profiles are shown in Fig.~\ref{stress_z}.  The top two plots show the Maxwell stress (solid line) and Reynolds stress (dashed line) for the two radial locations and as a function of $\beta_0$ using the same color scheme as before.  As with the magnetic energy, the dip in stress near the mid-plane is a result of the low ionization (Am = 1) region there. 

The bottom row shows the total Maxwell stress (solid line) and the {\it turbulent} component to this stress (dotted line) calculated by subtracting $-\langle B_x\rangle\langle B_y\rangle$ from $-\langle B_xB_y\rangle$.  With the exception of AD30AU1e3, the mid-plane region (i.e., within $|z| \lesssim 1-2H$) is completely dominated by turbulent stress.  Moving further away from the mid-plane, the large scale stress, $-\langle B_x\rangle\langle B_y\rangle$, plays an increasingly more important role, ranging from zero to $\sim 80$\% of the total stress in the active region.  In the case of AD30AU1e3, the stress is almost entirely large scale, with only a small region of turbulence at the mid-plane. 

The run AD30AU1e4L shows a stress profile very similar to AD30AU1e3, though at a lower amplitude at all $z$.  Furthermore, there is more turbulence near the mid-plane in AD30AU1e4L; within $|z| \lesssim 1.5H$, the turbulent component to the Maxwell stress is comparable to the total averaged stress.  

\begin{figure*}
\begin{minipage}[b]{10cm}
\begin{center}
\includegraphics[width=1.7\textwidth,angle=0]{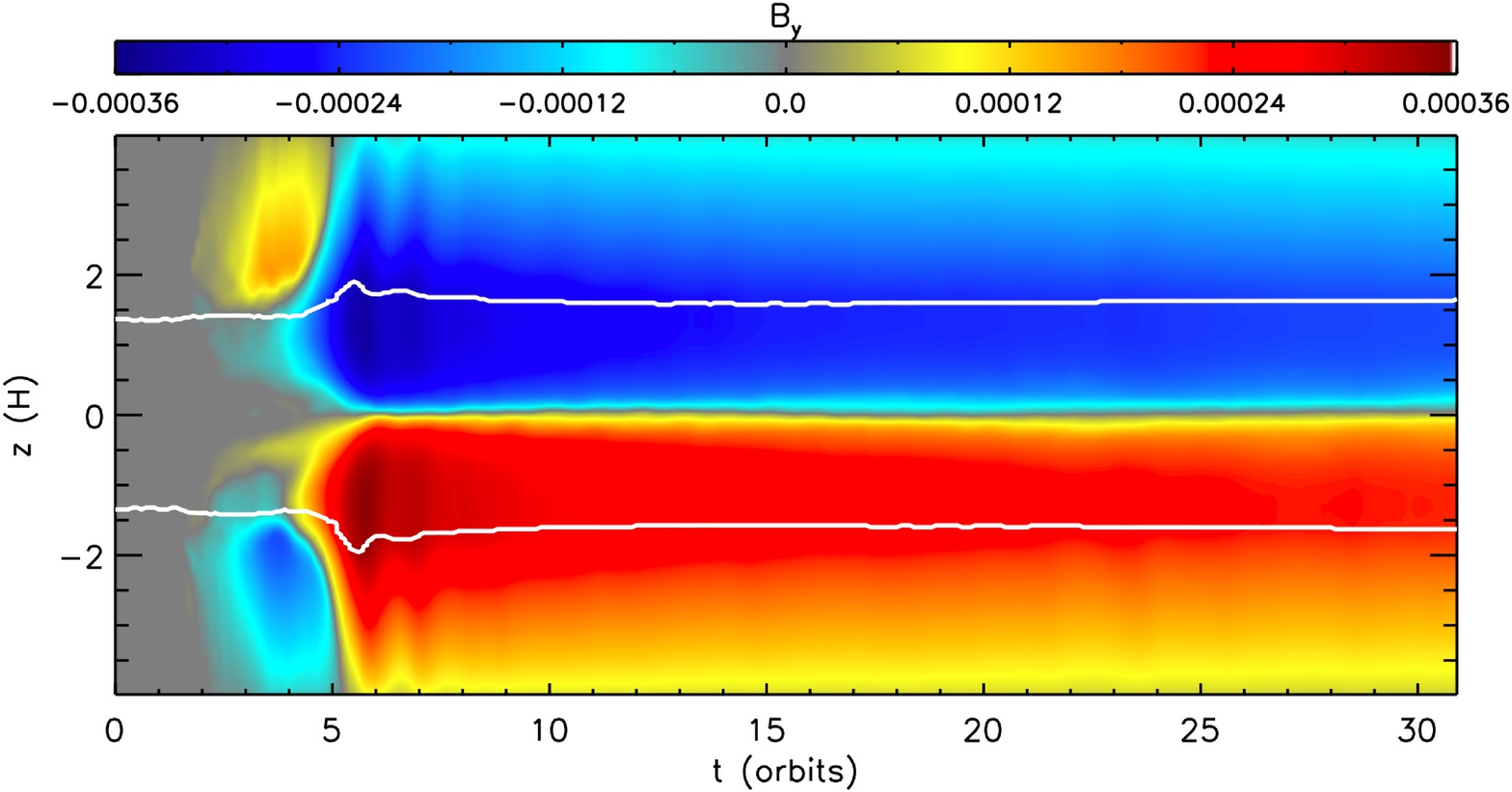}
\end{center}
\end{minipage}
\newline
\begin{minipage}[c]{10cm}
\begin{center}
\includegraphics[width=1.7\textwidth,angle=0]{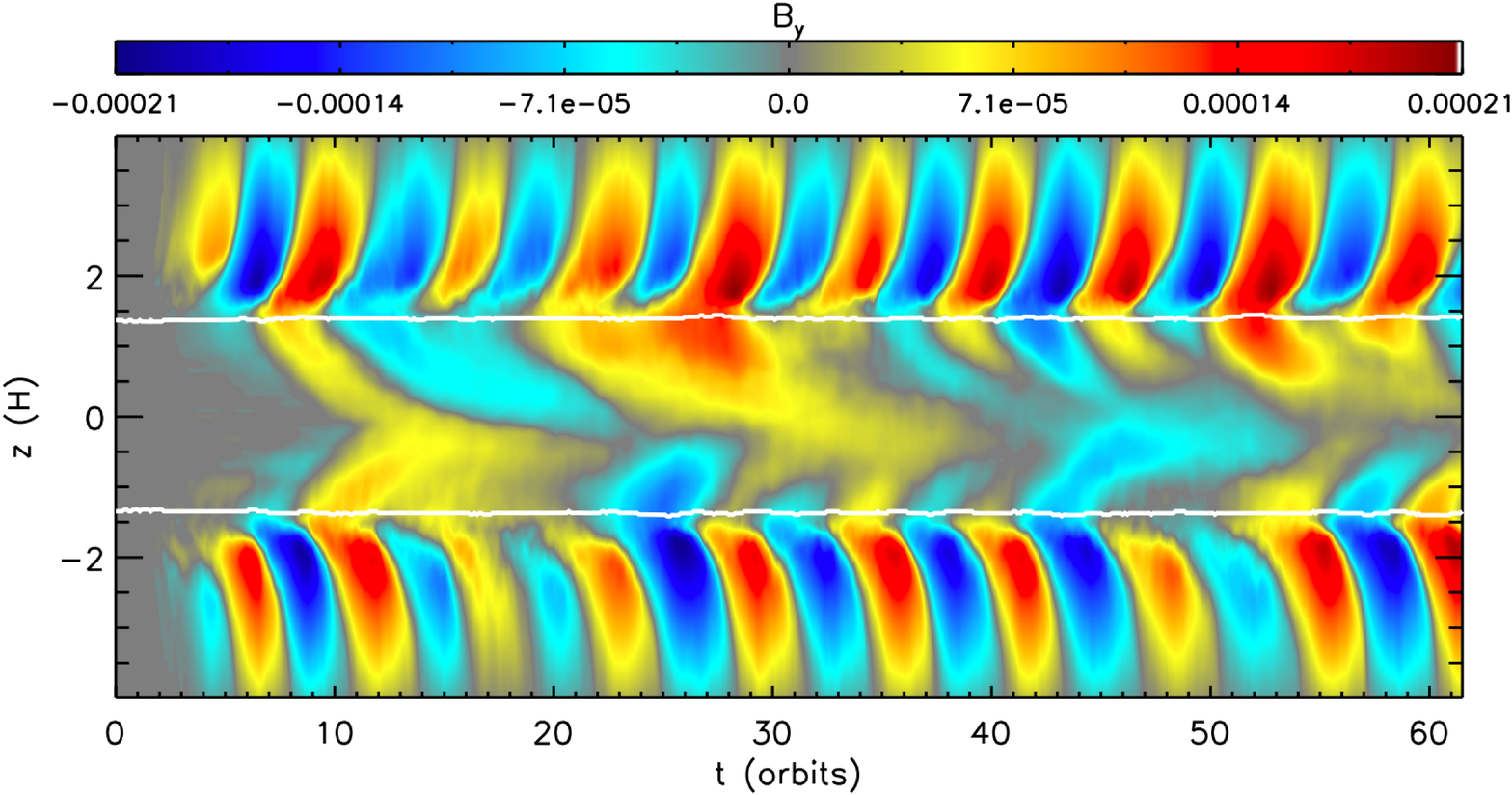}
\end{center}
\end{minipage}
\newline
\begin{minipage}[b]{10cm}
\begin{center}
\includegraphics[width=1.7\textwidth,angle=0]{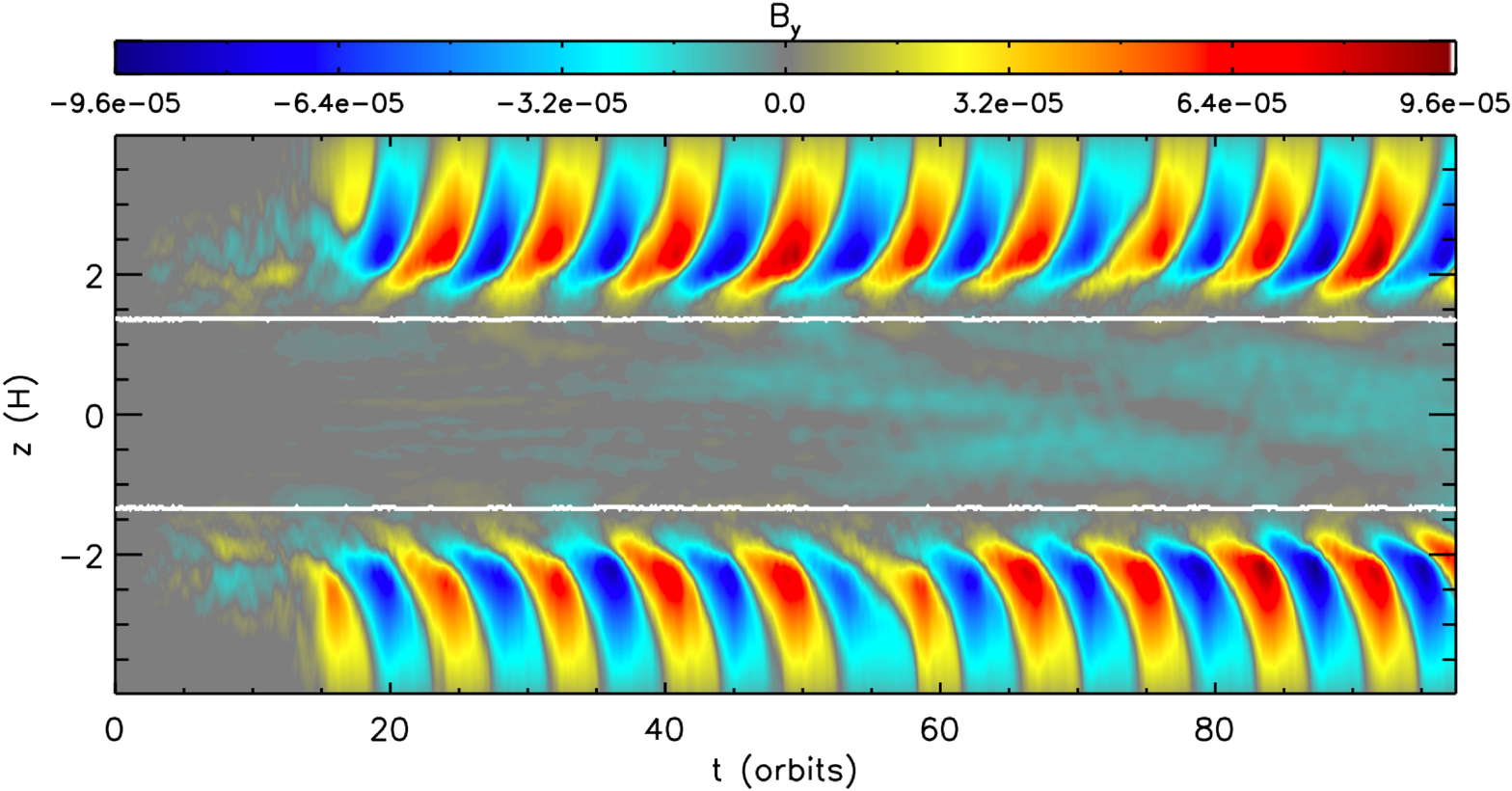}
\end{center}
\end{minipage}
\caption{
Space-time diagram of the horizontally-averaged toroidal field for AD30AU1e3 (top), AD30AU1e4 (middle), and AD30AU1e5 (bottom).  The white lines denote the base of the FUV ionization layers. The bottom two panels, (i.e., weaker field runs) show the MRI dynamo seen in many previous stratified MRI calculations; the toroidal field flips sign and buoyantly rises away from the mid-plane.  The period of this flipping is $\sim 6$ orbits for AD30AU1e4 and $\sim 9$ orbits for AD30AU1e5.  This pattern is strong for $|z| \gtrsim 2H$.  Closer to the mid-plane, the toroidal field behavior appears much less organized.  For the strong field case (top panel), the toroidal field eventually settles into a constant state with $B_y < 0$ for $z > 0$ and $B_y > 0$ for $z < 0$.  There is a strong current layer at $z = 0$.
}
\label{sttz_30au}
\end{figure*}

\begin{figure*}
\begin{center}
\includegraphics[width=\textwidth,angle=0]{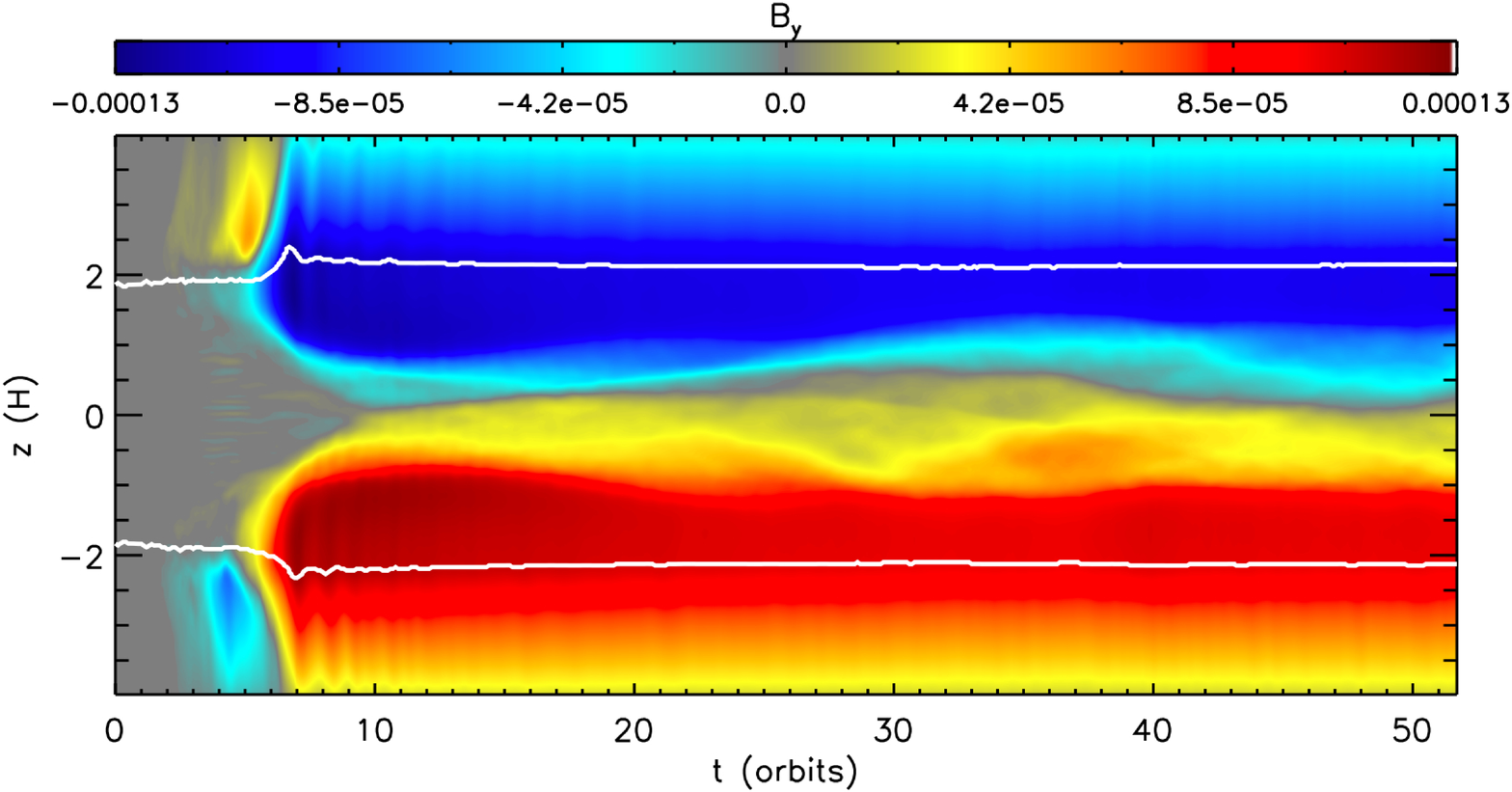}
\end{center}
\caption{
Space-time diagram of the horizontally-averaged toroidal field for AD30AU1e4L.  The white lines denote the base of the FUV ionization layers.  As with AD30AU1e3, the toroidal field settles into a constant state with $B_y < 0$ for $z > 0$, $B_y > 0$ for $z < 0$ and a strong current layer at $z = 0$.
}-
\label{sttz_sig0.01}
\end{figure*}

Another useful diagnostic is the space-time diagram of the horizontally averaged toroidal field, as shown in Fig.~\ref{sttz_30au} for the three simulations at 30 AU with $\Sigma_{\rm i} = 0.1$ g cm$^{-2}$ and in Fig.~\ref{sttz_sig0.01} for the $\beta_0 = 10^4$ simulation at this radius with $\Sigma_{\rm i} = 0.01$ g cm$^{-2}$.  In all of these plots, the white lines denote the locations of the FUV ionization front (i.e., where Am drops to unity).

From these figures, it is obvious that there is {\it some} activity in the Am = 1 region.  However, within this region, the field appears to be reduced in amplitude, in agreement with the dip in the time-averaged stress profiles.  With the exception of the top panel of Fig.~\ref{sttz_30au} and Fig.~\ref{sttz_sig0.01}, the familiar MRI dynamo behavior \cite[see, e.g.,][]{simon12} reemerges within the large Am regions.  This dynamo behavior manifests itself in the stress curves of Fig.~\ref{stress} as oscillations; we verified that the frequency of oscillation in these curves corresponds to the frequency seen in the $B_y$ dynamo pattern.  This effect points to the strong contribution of large scale correlations in $B_x$ and $B_y$ to the total Maxwell stress in the upper disk regions, as discussed above.

The runs AD30AU1e3 and AD30AU1e4L do not exhibit the MRI dynamo oscillations. The toroidal field remains stationary for the entire duration of the simulations, and changes sign across the disk mid-plane, leaving a thin layer with strong current. We will further examine this dichotomy in the next section.

\subsection{Quasi-laminar Flow vs Turbulence}
\label{results_lam}

As noted above, both runs AD30AU1e3 and AD30AU1e4L show different magnetic field behaviors (both in space and time) compared to the other simulations.
The difference between these two runs and the other simulations is further elucidated by considering the structure of the magnetic field. 

\begin{figure*}
\begin{center}
\includegraphics[width=\textwidth,angle=0]{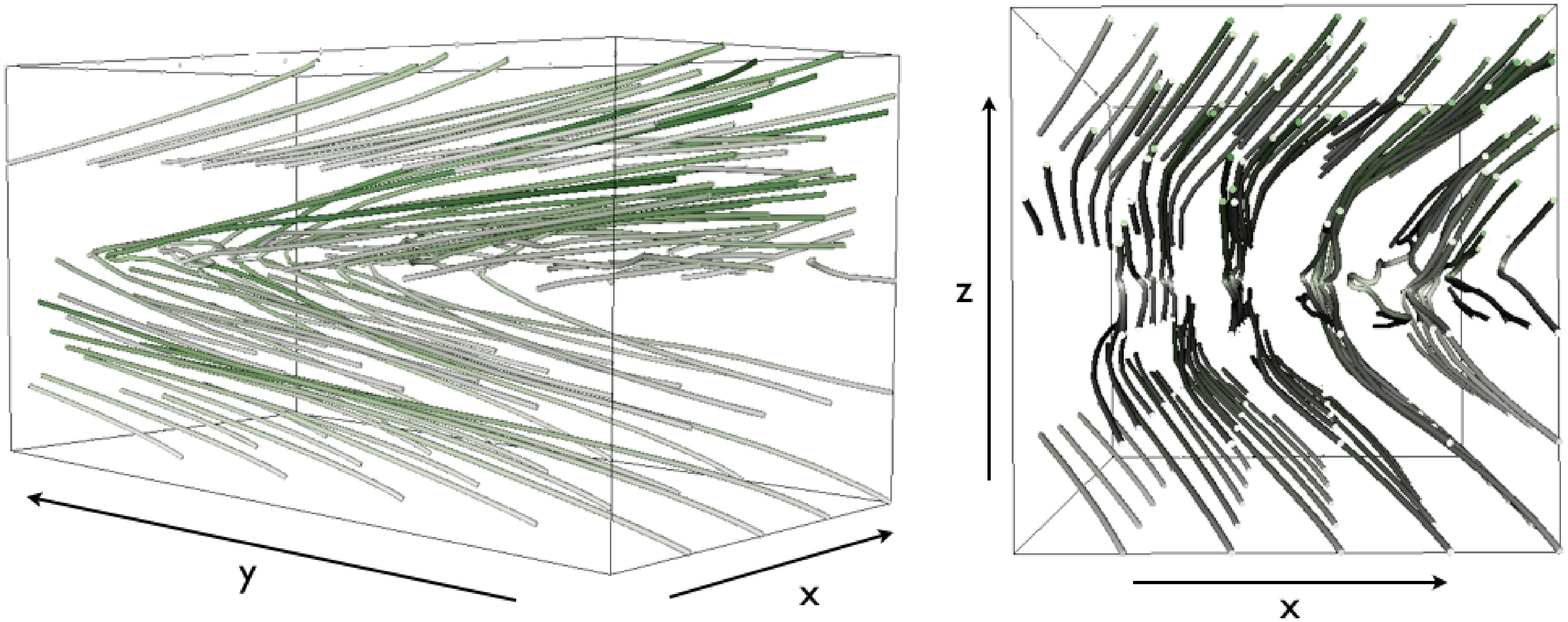}
\end{center}
\caption{
Rendering of magnetic field lines at orbit 20 in AD30AU1e3, shown from two different angles. The magnetic field structure is highly laminar, with a significant toroidal structure.  The field also has a structure consistent with a magnetocentrifugal wind and has an ``even-z symmetry" (see text).
}
\label{field_lines3}
\end{figure*}

\begin{figure*}
\begin{center}
\includegraphics[width=\textwidth,angle=0]{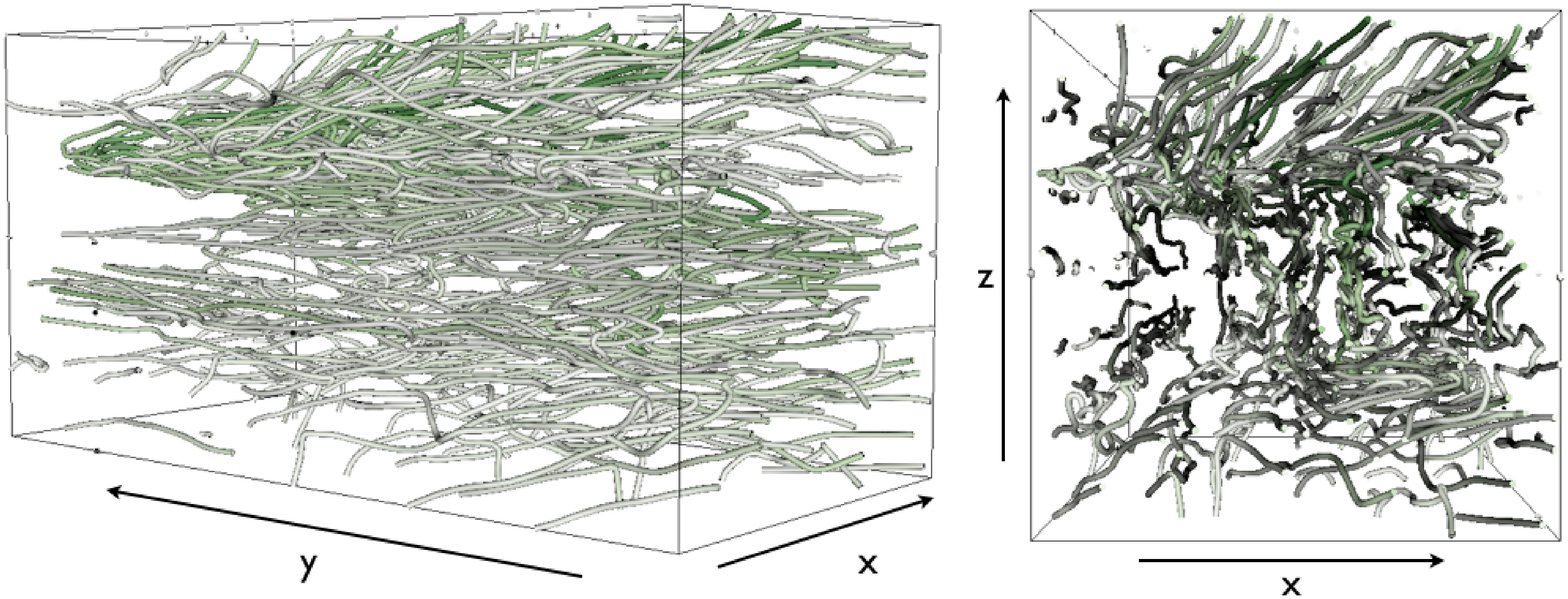}
\end{center}
\caption{
Rendering of magnetic field lines at orbit 20 in AD30AU1e4, shown from two different angles.  Unlike the structure in AD30AU1e3 (see Fig.~\ref{field_lines3}), the magnetic field is highly tangled in this run due to the turbulence present.  There does appear to be a significant toroidal component, as is usual for the MRI \cite[see, e.g.,][]{simon11a}.
}
\label{field_lines4}
\end{figure*}

Figures~\ref{field_lines3}~and~\ref{field_lines4} display volume renderings of magnetic field lines at one point in the saturated state of runs AD30AU1e3 and AD30AU1e4, respectively.  AD30AU1e3 has a largely laminar magnetic field structure, which is predominately toroidal.  In the poloidal plane (right panel of Fig.~\ref{field_lines3}), it is clear that a wind-like structure is present. AD30AU1e4 on the other hand has a very turbulent magnetic field structure, while still being predominantly toroidal.  This same turbulent structure is observed in AD30AU1e5 and AD30AU1e5L, whereas the quasi-laminar structure of AD30AU1e3 is also seen in AD30AU1e4L.

These results suggest that there are two classes of solutions here: one in which the flow is largely laminar, the other of which is turbulent. In the laminar cases, most of the $R\phi$ stress results not from small scale turbulent fluctuations, but large scale correlations in $B_x$ and $B_y$ as was shown in Fig.~\ref{stress_z}.   In the turbulent cases, there is a non-negligible fraction of the Maxwell stress in the FUV ionized region resulting from large scale correlations in the radial and toroidal fields.  In the Am = 1 region, the significantly weaker Maxwell stress appears to result from turbulence.

The reason for this dichotomy lies in the dependence of the MRI on the magnetic field strength in the ambipolar diffusion dominated regime. In unstratified simulations, \citet{bai11a} demonstrated that with strong ambipolar diffusion, the MRI operates when the magnetic field is sufficiently weak: increasing the net field will first lead to an increase of the turbulence level, until the field is too strong for the MRI to operate.
Above such threshold value of the field strength, the MRI tends to be suppressed and the flow transitions toward a more laminar state, with stress dominated by large-scale components. In the case of AD30AU1e3, the MRI is initially activated, but the {\it turbulent} $\beta$ pushes the system into a regime where ambipolar diffusion quenches the MRI. This is qualitatively similar to complete suppression of the MRI in the inner region of protoplanetary disks as studied by \citet{bai13b} and \cite{bai13c}, where both the effects of ambipolar diffusion and Ohmic diffusion were included. The main difference is that near the mid-plane, there is still weak turbulence (e.g., in the bottom left panel of Figure 5). This is because the ambipolar diffusion (rather than Ohmic resistivity) dominated mid-plane region is still linearly unstable to the MRI for moderate net vertical magnetic flux. 

Another related issue is why AD30AU1e4 shows strong MRI turbulence, whereas AD30AU1e4L does not.  AD30AU1e4L has a smaller ionization depth, which means that the region in which Am = 1 extends to larger $|z|$.  In the highly ionized layers, the total $\alf$ speed is too large for the MRI to operate, either based on the MRI instability criterion \citep{balbus98} or the result that when the field strength is sufficiently large, ambipolar diffusion quenches the MRI, as discussed in \cite{bai11a}.  AD30AU1e4, on the other hand, has a larger ionized layer (smaller Am = 1 region); this allows the $\alf$ speed to become small enough to permit the MRI in this region. In both of these runs, the Am = 1 region still remains weakly turbulent, consistent with our observation that the total $\alf$ speed and Am values in this region are the same in the two different cases.

\subsubsection{Wind Stress}
\label{results_wind}

As noted above, runs AD30AU1e3 and AD30AU1e4L both have a wind-like structure to the magnetic field with some turbulence within the Am = 1 region.  This magnetic field geometry can remove angular momentum from the disk via a Blandford-Payne type of mechanism \citep{blandford82,lesur13,bai13a,bai13b,fromang13}.  Furthermore, even in the fully turbulent systems, the presence of a net vertical field can lead to a non-zero $z\phi$ component to the stress tensor \citep{bai13a,fromang13}.  Following the arguments in \cite{fromang13}, we can integrate the angular momentum evolution equation over the disk height to define the $z\phi$ stress component,  

\begin{equation}
\label{wzphi}
W_{z\phi} \equiv \frac{(\rho v_z\delta v_y-B_zB_y)}{\rho_0\cs^2}\bigg|^{z_{\rm bw}}_{-z_{\rm bw}},
\end{equation}

\noindent
where $\pm z_{\rm bw}$ are the limits of the height integration.  The exact values for these limits will be discussed shortly.  To make this stress dimensionless, we normalize by the mid-plane gas density multiplied by the square of the sound speed. All of the stresses that we have considered thus far have only been due to the $R\phi$ component of the stress tensor.  This term thus represents an {\it additional} source of angular momentum transport.   We must note here, however, that because we are normalizing by mid-plane quantities in the definition of $W_{z\phi}$ but by a volume average in the definition of $W_{R\phi}$, the values for the two different stresses cannot be compared directly to one another.

The local shearing-box approximation has the limitation that the radially inner and outer sides of the box are symmetric because curvature is ignored. This leads to two possible types of outflow geometry, depending on whether the outflow from the top and bottom sides of the box flow toward the same or opposite radial directions. The former case is termed as ``even-$z$" symmetry, while the latter case is termed as ``odd-$z$" symmetry (see Section 4.4 of \citet{bai13b} for a thorough description). The desired (physical) outflow geometry follows the ``even-$z$" symmetry. In the study of the inner region (1 AU) of protoplanetary disks, \citet{bai13b} found that both geometries are equally possible in their shearing-box simulations which yielded pure laminar wind, and argued that real systems should proceed in the physical manner. However, in the ideal MHD simulations of \citet{bai13a}, it was found that the ``odd-$z$" symmetry prevails when stress is dominated by large-scale magnetic field. Moreover, they showed that the presence of the MRI dynamo makes the radial direction of the outflow change sign in time. We observe this feature in our non-laminar simulations as well, which argues against the outflow resulting from a physical disk wind. Clearly, the issue with symmetry needs to be resolved with global simulations. For our study, we set aside this issue and just focus on other aspects of the outflow. In particular, we assume (with these caveats in mind) that the outflow {\it always} follows the physical ``even-z symmetry" wind discussed above, which gives 

\begin{equation}
\label{wzphi_symm}
W_{z\phi}\bigg|_{z_{\rm bw}} = -W_{z\phi}\bigg|_{-z_{\rm bw}}.
\end{equation}

\noindent
Thus, the quantity of interest for our simulations will be $2 |W_{z\phi}|_{z_{\rm bw}}$ (the factor of 2 resulting from the assumption made in equation~(\ref{wzphi_symm})). 

Following the works of \cite{wardle93} and \cite{bai13a}, the location for the base of the wind $z_{\rm bw}$ is defined as the location where the flow transitions from sub-Keplerian to super-Keplerian, which translates to the location where $\delta v_y$ changes sign. In practice, this is conveniently determined for laminar flow. In the presence of MRI turbulence, the horizontally averaged value of $\delta v_y$ varies and changes sign due to the MRI dynamo. To properly define $z_{\rm bw}$ in this case, we calculate the vertical profile for the absolute value of $\delta v_y$, and average it over time, giving us $\overline{|\delta v_y|}$. The location of $z_{\rm bw}$ is found by determining the first local minimum $\overline{|\delta v_y|}$ in $z$ as one moves towards the mid-plane from either vertical boundary.  The result is two values for $z_{\rm bw}$, one on either side of the mid-plane.  For simplicity, we average these two numbers to arrive at a final value for $z_{\rm bw}$.  

\begin{figure*}
\begin{minipage}[!ht]{8cm}
\begin{center}
\includegraphics[width=1\textwidth,angle=0]{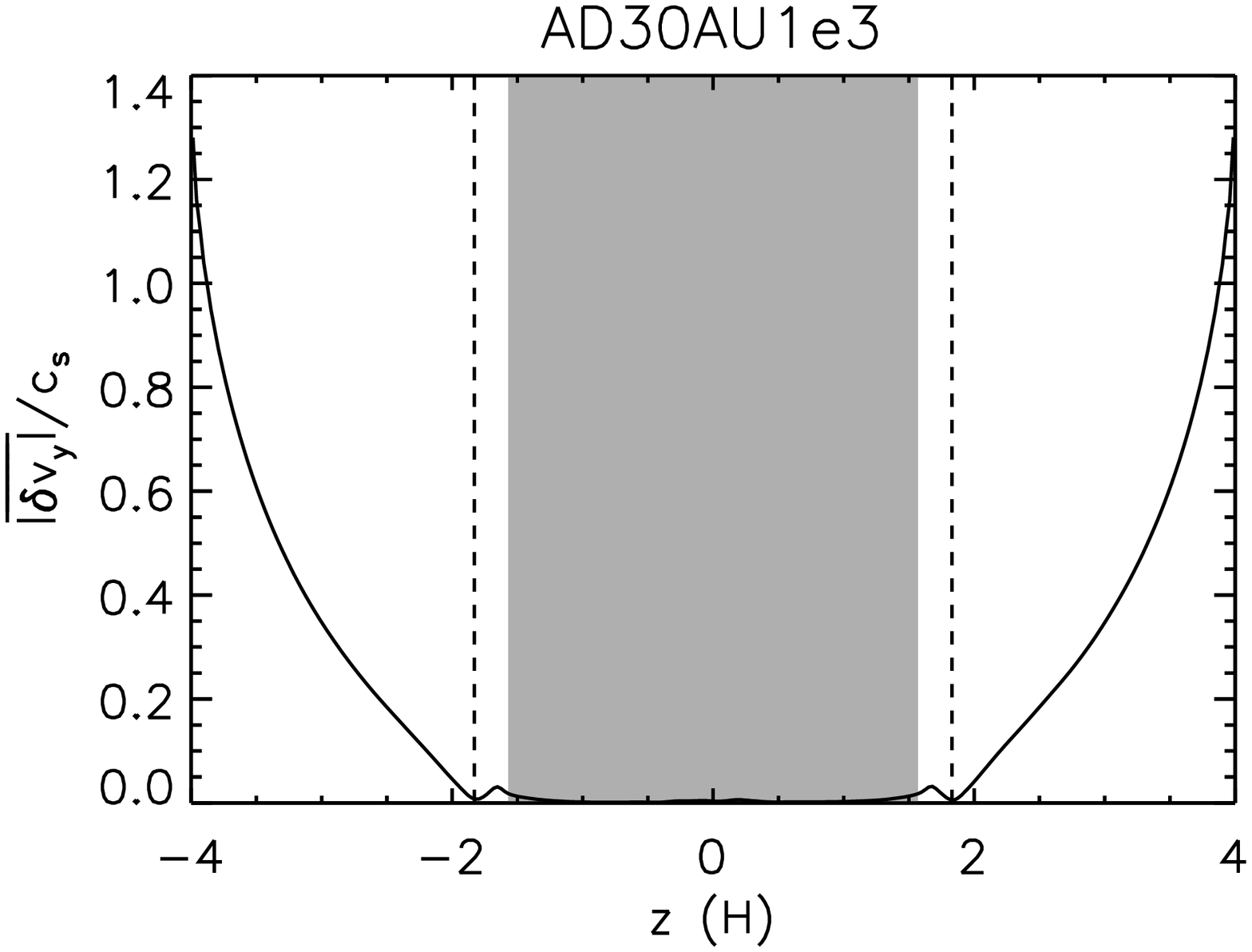}
\end{center}
\end{minipage}
\begin{minipage}[!ht]{8cm}
\begin{center}
\includegraphics[width=1\textwidth,angle=0]{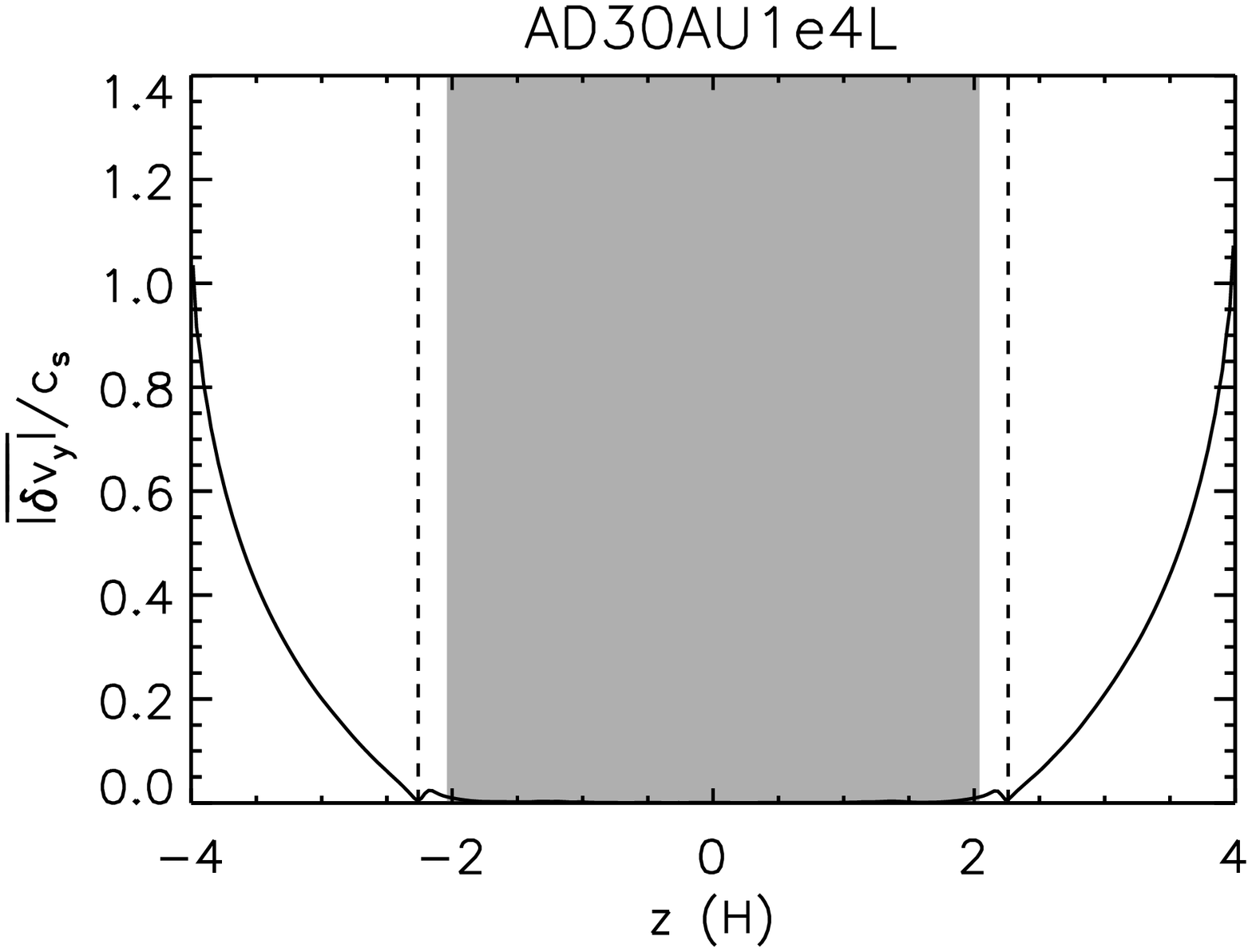}
\end{center}
\end{minipage}
\caption{
Vertical profile for $\overline{|\delta v_y|}$ as defined in the text, normalized by the gas sound speed for AD30AU1e3 (left) and AD30AU1e4L (right).  In both runs, $\overline{|\delta v_y|}$ rapidly increases beyond the base of the wind as defined in the text. The locations of the wind base are denoted by the vertical lines. The Am = 1 region here (shaded in gray) is slightly larger than the one in the upper left panel of Fig.~\ref{stress_z} because we averaged the $z$ values for the edges of the Am = 1 region across the multiple simulations plotted in Fig.~\ref{stress_z} as described in that section.
}
\label{vy_wind_laminar}
\end{figure*}

This approach gives the canonical location of $z_{\rm bw}$ for the quasi-laminar flows; see Fig.~\ref{vy_wind_laminar}, which shows the vertical profile of $\overline{|\delta v_y|}$ for the quasi-laminar runs. In the more turbulent field geometries, the physical motivation for this definition of $z_{\rm bw}$ is a little less clear.  While we are able to determine $z_{\rm bw}$ for the turbulent runs using the method described above, we must also keep in mind that there is a non-negligible turbulent component to the stress (see Fig.~\ref{stress_z}) for $|z| > z_{\rm bw}$.   In other words, this turbulence can still contribute to mass accretion above the base of the wind.  This motivates us to consider $z_{\rm bw} = 4H$ (the vertical boundaries of the domain) as an alternative choice for the limits of integration. 

We use these two different definitions of $z_{\rm bw}$ to calculate both an $R\phi$ stress and a $z\phi$ stress (see Table~\ref{tbl:sims}).  $\alpha_{\rm bw}$ is the $R\phi$ stress integrated from $-z_{\rm bw}$ to $+z_{\rm bw}$, where $z_{\rm bw}$ is defined by the $\overline{|\delta v_y|}$ criterion. Similarly, $2\overline{|W_{z\phi}|}_{\rm bw}$ is the $z\phi$ stress evaluated at $z_{\rm bw}$, again using this same criterion for $z_{\rm bw}$. The subscript ``total" in the table equates to choosing $z_{\rm bw} = 4H$; thus, $\alpha_{\rm total}$ ($2\overline{|W_{z\phi}|}_{\rm total}$) is the $R\phi$ ($z\phi$) stress calculated for the entire vertical domain.

\subsubsection{The Ambipolar Damping Zone}
\label{results_damp}

In Paper I, where no net vertical magnetic flux was included, we found that the Am = 1 region corresponded to no MRI turbulence.  The only stress that was present in this region was Reynolds stresses induced by the active layers and Maxwell stress resulting from large scale correlations in the radial and toroidal magnetic fields.  Thus, we referred to this region as the ambipolar dead zone.  However, in the presence of net vertical magnetic flux, the Am = 1 region is still MRI unstable, and will result in turbulence albeit at lowered saturation amplitude.  Indeed, we see from Fig.~\ref{stress_z} that turbulent stresses are present within the Am = 1 region.  For the quasi-laminar field geometries, we see a strong Maxwell stress in the Am = 1 region, and an appreciable Reynolds stress.

While the MRI is active in the Am = 1 region, the stress values here are lowered compared to the peaks of the stress in the FUV active regions; the MRI is damped, but not dead. In the case of approaching zero net vertical flux, $\beta_0 \rightarrow \infty$, this damped region becomes the ``dead region" of Paper I. 

In order to generalize the effect of the Am = 1 region on the MRI, we now refer to it as the {\it ambipolar damping zone}.   Accretion through this region is possible via different mechanisms that depend on the strength of the background vertical field; for strong field, the stresses in this region are a combination of strong laminar and turbulent stresses, while for weaker fields, these stresses are turbulent and saturate at a lower amplitude.  For zero vertical field, accretion still proceeds through the presence of Reynolds stresses induced by the active layers and weak large scale correlations in the radial and toroidal magnetic fields.

\subsection{Low Ionization Depth Simulations}
\label{results_lowion}

In this section, we explain the similarity between the stress amplitudes in the lower ionization runs (AD30AU1e4L, AD30AU1e5L) and their higher ionization counterparts (AD30AU1e4, AD30AU1e5) as shown in Fig.~\ref{stress}.

The small difference between the stress evolution of AD30AU1e5 and AD30AU1e5L is a result of the resolution employed in these simulations.  As mentioned above for the ideal MHD runs, in the region where $Q_z$ is relatively small, the temporally and horizontally averaged stress is reduced.  When viewing the equivalent of Fig.~\ref{qz_z_initial} for AD30AU1e5L, we found that the under-resolved region spans the range $-2H < z < 2H$.  Within this region, the stress is slightly damped due to the lower resolution there.  Thus, in reality, the stress level for AD30AU1e5 should be somewhat higher than what is shown in the figure.  We cannot quantify the exact value of the converged stress level, however, given the computational expense of running a higher resolution simulation at the large domain size needed for these non-ideal MHD calculations. In AD30AU1e5L, the ambipolar damping zone is roughly equivalent to the under-resolved region, and so the stress level for this run is more accurate.  The bottom line is that AD30AU1e5 (and by the same argument AD100AU1e5) will, in reality, have a higher stress value than what is shown in the figure and Table~\ref{tbl:sims}.  Thus, these values should not be taken as exact, but as an approximate, order of magnitude estimate for the true stress values; for the purposes of this paper, this is sufficient.

Because the region in which the MRI is not as well resolved is smaller for the $\beta_0 = 10^4$ simulations than in the $\beta_0 = 10^5$ simulations, the above argument does not necessarily apply to AD30AU1e4 and AD30AU1e4L.  We are not entirely sure why the stress levels are similar between these two calculations.  One possibility is that due to the quasi-laminar wind-like nature of AD30AU1e4L, the stress in the ambipolar damping zone is not as drastically different from the peak stress values (near $|z| \sim 2H$) as in AD30AU1e4.   This feature can also be seen in Fig.~\ref{stress_z} for AD30AU1e3; the peak to trough stress ratio is much smaller for the laminar run vs. the turbulent runs.  We see a similar effect in AD30AU1e4L, which means that while the entire vertical stress profile is lower in amplitude in AD30AU1e4L compared to AD30AU1e4, the ambipolar damping zone in AD30AU1e4L has a larger stress than it would if this run were instead turbulent.  This effect may very well equate to the height-averaged stress values being similar in the two $\beta_0 = 10^4$ runs.

\section{Implications}
\label{discussion}

\subsection{Mass Accretion Rates}
\label{mass_accretion_rates}

As was done in Paper I, we estimate an accretion rate corresponding to the stresses observed in our simulations.  Following Section 4.1 of Paper I, the mass accretion rate due to the $R\phi$ stress, assuming a steady state, is

\begin{equation}
\label{mdot_rphi}
\dot{M}_{R\phi} = \frac{2\pi\Sigma\cs^2}{\Omega}\alpha.
\end{equation}

\noindent
where, for the purposes of deriving the accretion rates, we drop any subscripts on $\alpha$.  Equation~(\ref{mdot_rphi}) was derived assuming that the only contribution to accretion is $W_{R\phi}$ and then solving the angular momentum conservation equation.  If we now assume that the only stress contributing to accretion is $W_{z\phi}$, we can solve the angular momentum equation, again assuming a steady state, to obtain an equivalent accretion rate due to this stress,

\begin{equation}
\label{mdot_zphi}
\dot{M}_{z\phi} = \frac{8\sqrt{\pi} R \Sigma \cs^2}{\Omega H}\overline{|W_{z\phi}|}.
\end{equation}

\noindent
where again, we drop any subscripts for $\overline{|W_{z\phi}|}$.  

An estimate for the total accretion rate is acquired by simply adding equations~(\ref{mdot_rphi}) and (\ref{mdot_zphi}),  

\begin{equation}
\label{mdot_tot}
\dot{M} = \frac{2\pi\Sigma\cs^2}{\Omega}\left[\alpha + \frac{4}{\sqrt{\pi}}\frac{R}{H}\overline{|W_{z\phi}|}\right].
\end{equation}

\noindent
Note that the first (second) term of equation~(\ref{mdot_tot}) was derived assuming that $W_{R\phi}$ ($W_{z\phi}$) is the only stress contributing to mass accretion. This derivation is obviously not self-consistent. A more rigorous derivation \cite[e.g.,][]{fromang13} would require assumptions of various radial gradients in the disk that depend on global disk structure (which are also uncertain). However, for our purposes, the above formula is sufficient to give order-of-magnitude estimates of accretion rates.  

From Table~\ref{tbl:sims}, we estimate that $\dot{M} \sim 10^{-7}$--$10^{-6} M_{\sun}/{\rm yr}$ for $\beta_0 = 10^3$, $\dot{M} \sim 10^{-8}$--$10^{-7} M_{\sun}/{\rm yr}$ for $\beta_0 = 10^4$, and $\dot{M} \sim 10^{-8} M_{\sun}/{\rm yr}$ for $\beta_0 = 10^5$.  The $z\phi$ component of the stress accounts for a non-negligible fraction of these accretion rates.  Indeed, if we consider only the turbulent runs (and the case where $z_{\rm bw} = 4H$), we find that $\dot{M} \sim 10^{-8} M_{\sun}/{\rm yr}$ for $\beta_0 = 10^4$ and $\dot{M} \sim 10^{-9}$--$10^{-8} M_{\sun}/{\rm yr}$ for $\beta_0 = 10^5$; the accretion rates are lower when the $z\phi$ stress is ignored.  This result is not surprising since the contribution of the $z\phi$ stress to $\dot{M}$ dominates by a factor of roughly $R/H$ for comparable values of the $R\phi$ and $z\phi$ stresses (see Equation~(\ref{mdot_tot})).

Despite the relative importance of the $z\phi$ stress, we must be cautious in our interpretation of this result.   As described above, the MRI dynamo leads to a temporally oscillatory behavior of $W_{z\phi}$ around zero. The stress values in Table~\ref{tbl:sims} are the time-average of the {\it absolute value} of $W_{z\phi}$ while in reality it may average to zero. Based on the same argument of \citet{bai13a} for their ideal MHD simulations, we tentatively suggest that the outflows observed in the turbulent simulations may not efficiently carry away disk angular momentum; disk accretion would then be mostly driven by radial transport ($W_{R\phi}$). 

Regardless of whether or not we ignore the $z\phi$ stress, our accretion rate estimates are in good agreement with observational constraints \cite[e.g.,][]{gullbring98,hartmann98a} for $\beta_0 = 10^4$--$10^5$; the strongest field case, $\beta_0 = 10^3$, produces accretion rates that are too large when compared with these constraints.    Thus, relatively weak background vertical fields, with $\beta_0$ values of $10^4$--$10^5$, are the best configuration to produce expected accretion rates.  These $\beta_0$ values correspond to vertical field strengths of $\sim 60$--200~$\mu$G and 10--30~$\mu$G at 30 AU and 100 AU, respectively.

Our results naturally connect to recent studies of \citet{bai13b} and \citet{bai13c} that focused on inner regions of protoplanetary disks. Both Ohmic resistivity and ambipolar diffusion were included in their simulations as appropriate for the inner disk. They found that first of all, net vertical magnetic flux is essential to drive rapid accretion consistent with observations. In addition, the MRI can be completely suppressed in the inner disk up to about 10 AU, with accretion purely driven by disk wind under the physical wind geometry. Our simulations in Paper I and this paper focus on the ambipolar diffusion dominated outer disk, with a similar conclusion that net vertical magnetic flux is essential. However, unlike the inner disk regions, the MRI is likely the dominant driving mechanism of accretion in the outer disk.

Finally, it is worth noting that our results are in general agreement with the work of \cite{salmeron07}, in which semi-analytic methods were employed to study angular momentum transport from both the MRI {\it and} a magnetic wind.  Vertical fields can potentially transport angular momentum via a wind in addition to the MRI, and the nature of the accretion flow depends on the strength of the vertical field.

\subsection{Disk Structure and Stress Scale}
\label{structure}

One advantage to studying the MRI using local simulations is that MRI turbulence is itself quite local (i.e., turbulent fluctuations are on scale $\lesssim H$) \citep[e.g.,][]{guan09a,simon12}.   However, in these simulations, either the nature of the turbulent magnetic fields lead to $R\phi$ stresses that have a non-negligible large scale component or the magnetic field is mostly large scale and laminar (as in the runs AD30AU1e3 and AD30AU1e4L).  In the calculations where turbulent transport is still present, this turbulence becomes less important compared to the large-scale stress further away from the mid-plane (see Fig.~\ref{stress_z}).  Furthermore, the reasonably strong accretion rates due to the vertical field torques point to the importance of large scale vertical fields threading the disk and transporting angular momentum away via a wind.

These considerations imply that global simulations may be better tools for understanding the structure and evolution of protoplanetary disks, at least in regions where ambipolar diffusion becomes important.  Indeed, \cite{gammie98b} suggests that if structures relevant to accretion exist on scales $>> H$, then the system can no longer be considered ``local".   However, if the stress values ($W_{R\phi}$ and $W_{z\phi}$) are determined solely by {\it local} quantities, then using local simulations may still be sufficient.   A comparison between local and global simulations (not unlike the work of \cite{sorathia12}, \cite{beckwith11}, and \cite{simon12}) in the context of protoplanetary disks may be a fruitful avenue to pursue in order to address these issues.

\begin{figure*}
\begin{minipage}[!ht]{8cm}
\begin{center}
\includegraphics[width=1\textwidth,angle=0]{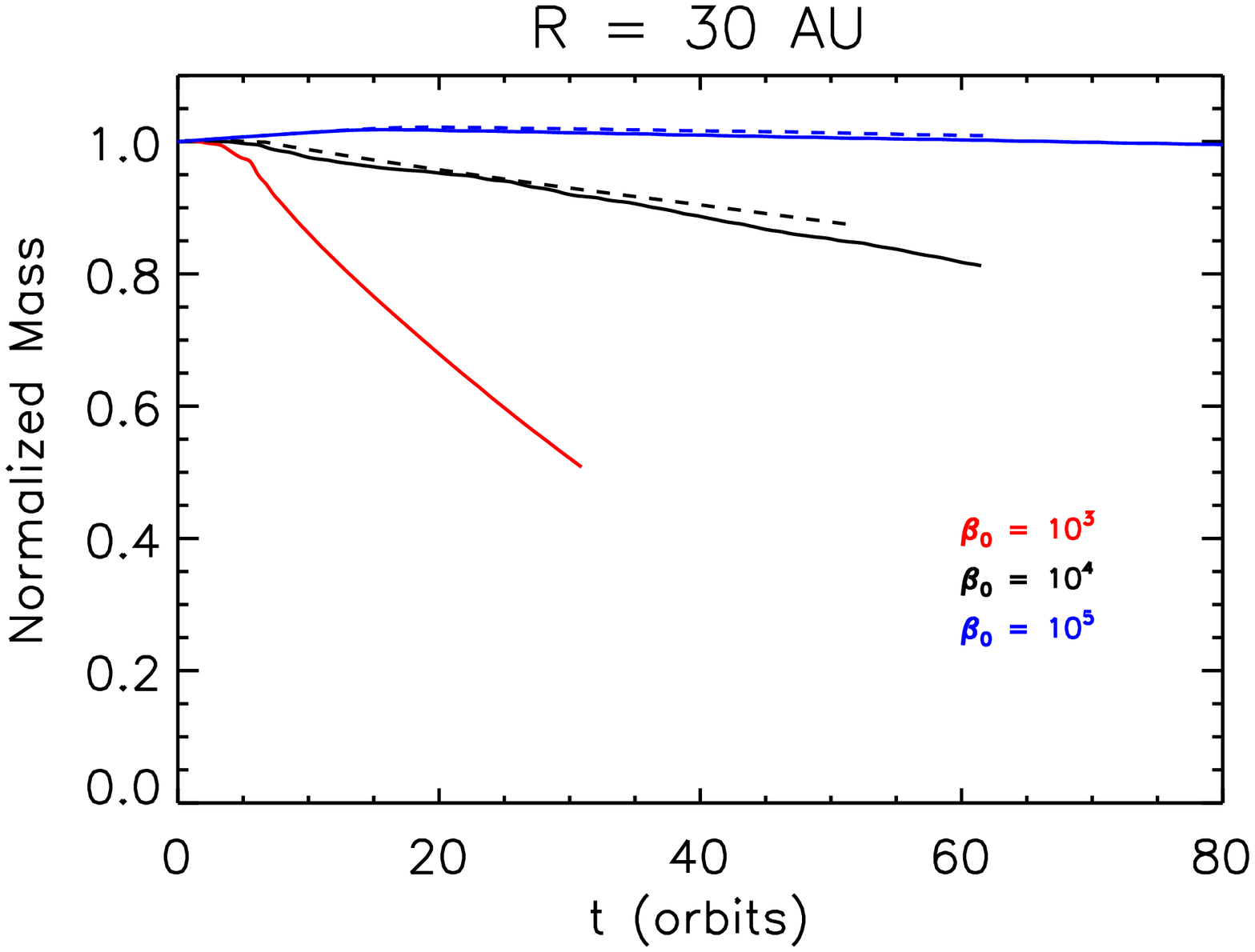}
\end{center}
\end{minipage}
\begin{minipage}[!ht]{8cm}
\begin{center}
\includegraphics[width=1\textwidth,angle=0]{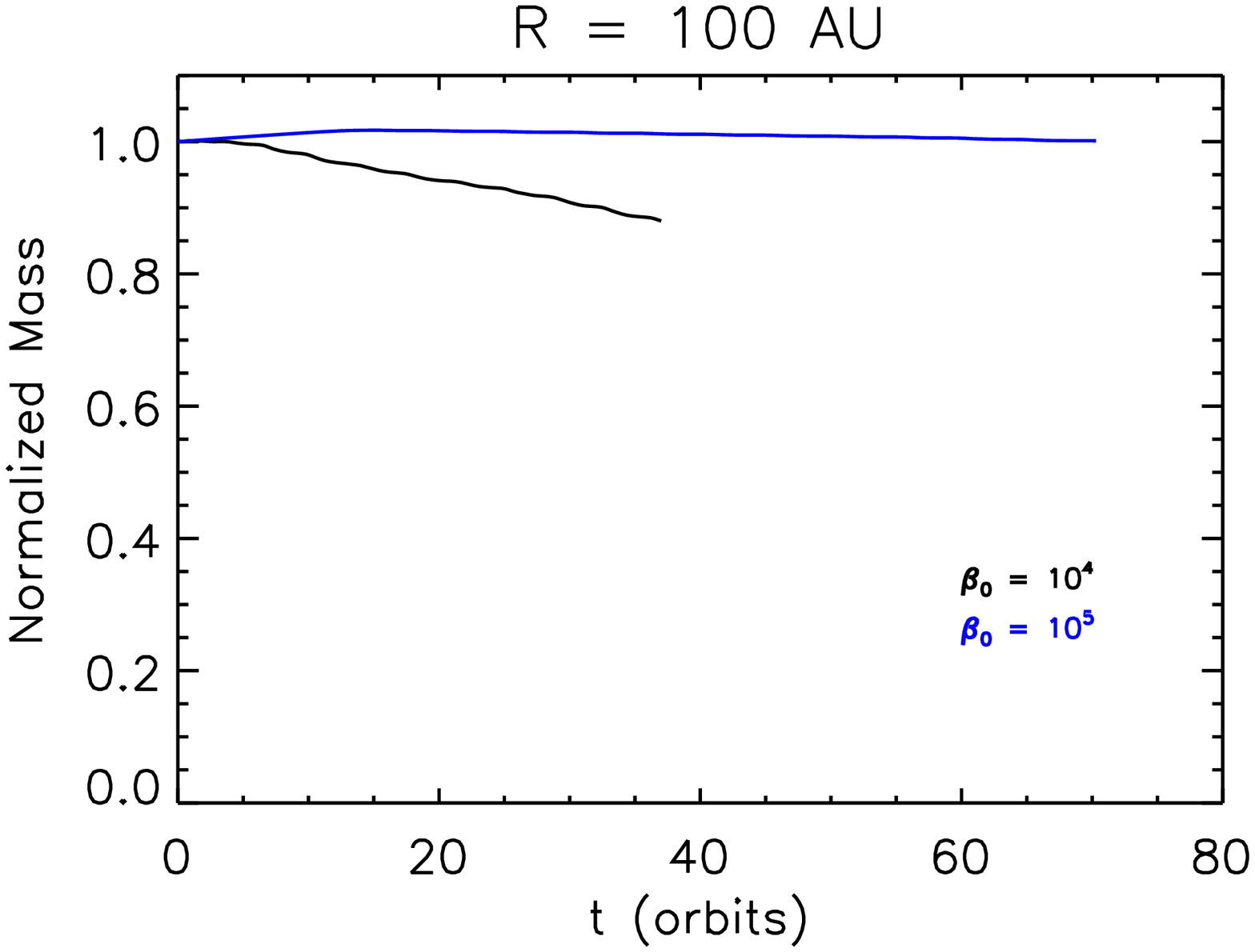}
\end{center}
\end{minipage}
\caption{
Total mass versus time (normalized by the total initial mass) for the non-ideal MHD simulations with a layered Am profile. The left panel corresponds to the runs at 30 AU and the right panel corresponds to 100 AU.  As described by the legend, the color of the line represents the strength of the net vertical magnetic field; the red line corresponds to $\beta_0 = 10^3$, the black lines correspond to $\beta_0 = 10^4$, and the blue lines correspond to $\beta_0 = 10^5$.  The solid lines correspond to an ionization depth of $\Sigma_{\rm i} = 0.1$ g cm$^{-2}$, and the dashed lines correspond to $\Sigma_{\rm i} = 0.01$ g cm$^{-2}$.   For $\beta_0 \lesssim 10^4$, there is significant mass lost from the domain, and the mass loss rate increases with decreasing $\beta$.
}
\label{mass}
\end{figure*}

\subsection{Mass Loss}
\label{mass_loss}

Our simulations show a significant rate of mass loss resulting from a combination of the disk wind and turbulence. Figure~\ref{mass} shows the time evolution of the mass in our layered Am simulations.  For the weakest field case, there is no mass loss. However, as $\beta_0$ is decreased, the mass loss rate grows; at $\beta_0= 10^3$ and 30 AU, nearly half of the mass has been lost by orbit 30.  This behavior has been observed in other shearing box simulations incorporating a vertical net field geometry \citep{suzuki09,bai13a,lesur13,fromang13}.

While this observed behavior is likely a physical result due to the launching of a magnetic wind (and perhaps turbulent motions as well), we cannot accurately quantify the mass loss rates in our local simulations.  \cite{fromang13} showed that as the vertical domain size is increased, the mass loss rate from the domain is reduced significantly.  In particular, they found that as the vertical height is increased, the point at which the flow velocity equals the fast magnetosonic speed always lies at the edge of the vertical boundary.  These considerations and similar arguments made by \cite{bai13a,bai13b} and \cite{lesur13} suggest that global simulations are necessary to capture an accurate representation of vertical disk mass loss. 

With that critical caveat, however, it is clear that the mass loss rates seen here could be important for protoplanetary disk dispersal 
even if they are substantially overestimated. Observational estimates for the time scale of final disk dispersal are of the order of 
$10^5 \ {\rm yr}$ \cite[e.g.,][]{wolk96}, about 10\% of the disk lifetime or $\sim 100-1000$ orbits in the region 
between 30~AU and 100~AU. Disk dispersal is typically attributed to thermal disk winds (``photoevaporation"), that may be driven by ionizing extreme-ultraviolet (EUV) photons, stellar X-rays, or the same FUV photons important here for ionization \citep{hollenbach94,alexander06,gorti09,owen11}. Our results imply that weak net fields are necessary for accretion, and hint that those same fields may result in MHD mass loss rates from the outer disk that are comparable to or (when compared against EUV mass loss rates) larger than those from photoevaporation. The fact that the rate of mass loss {\em increases} strongly as $\beta_0$ drops suggests an evolutionary scenario in which the final dispersal of the disk occurs at an accelerating pace as mass is lost (assuming net magnetic flux is conserved). A combination of evolutionary models and global disk simulations are needed to investigate such suggestions.

\section{Conclusions}
\label{conclusions}

We have carried out a series of stratified shearing box simulations in the presence of a net vertical field.  We have run both ideal MHD simulations (in order to study convergence of the turbulent saturation level with resolution) and non-ideal MHD simulations that include ambipolar diffusion profiles relevant to the outer regions of protoplanetary disks.  This work serves as a companion to our previous work, \cite{simon13a}, in which we showed that ambipolar diffusion significantly damps the MRI in the absence of a net vertical field. Our conclusions are as follows,

\begin{itemize}

\item Our simulations are reasonably well-converged at a resolution of 36 grid zones per $H$.  The regions on the grid where characteristic vertical MRI modes are not well-resolved do not play a significant role in the total turbulent saturation level.

\item We observe the similar layered structure of accretion as was observed in Paper I; there is a region of reduced stresses near the disk mid-plane surrounded by FUV-ionized layers on either side where the stress values are larger.  The stresses near the mid-plane are reduced by roughly an order of magnitude compared to the peak values (with the exception of the strongest vertical field case, for which the reduction is less severe).  We refer to this region as the {\it ambipolar damping zone}.

\item The presence of a net vertical magnetic field greatly enhances the total stress level compared with the absence of such a field. The saturation level of the stress increases rapidly as the net vertical field becomes stronger relative to the gas pressure. Accretion rates estimated from the stresses lie within the range of values expected from observations for background vertical field strengths of $\beta_0 = 10^4$--$10^5$.  These field strengths correspond to $\sim 60$--200~$\mu$G and 10--30~$\mu$G at 30 AU and 100 AU, respectively.

\item The solution becomes largely laminar at the low-$\beta$ surface layer of the disk, with a large scale $R\phi$ stress as well as a magnetocentrifugal type wind. Toward the disk mid-plane, the $R\phi$ stress is primarily in the form of turbulent fluctuations unless the net vertical magnetic flux is sufficiently large.

\item There is significant outflow from the simulations in the presence of net vertical magnetic flux, and the mass loss rate increases strongly with increasing net vertical magnetic flux. 

\end{itemize}

We are aware of the limitations of using the shearing-box approximation to study accretion disks in the presence of an outflow, particularly in how the large-scale stress and outflow properties could depend on the vertical size of the box \citep{fromang13,bai13b}, thus invoking the need for global simulations.
Due to this uncertainty, the exact values for our estimated stress levels and accretion rates should be taken cautiously.  We are, however, confident in our result that the presence of a net vertical field greatly enhances the stress levels, and we therefore believe our estimates of the accretion rates to be correct to within an order of magnitude or so.  

It may also be instructive to apply different disk models (other than the MMSN) to our machinery, thus exploring the sensitivity of our results to the underlying assumptions for the ionization and density structure.

One final uncertainty lies in neglecting the Hall effect \cite[see, e.g.,][]{Wardle99}, which may very well be important at the radii we have considered here \citep{armitage11}. The linear regime of the MRI in the presence of the Hall term was explored by \cite{Wardle99}, \cite{balbus01},
and \cite{wardle12}. These authors showed that the MRI growth rate is strongly affected by the sign of ${\bmath \Omega} \cdot {\bmath B}$, i.e., how
the vertical magnetic field is aligned with the angular velocity vector.   It remains to be quantified how the Hall effect will influence the MRI in the non-linear regime, particularly at the large radial distances considered in this work, where ambipolar diffusion is significant. 

These uncertainties aside, our results show that the presence of some non-zero field component perpendicular to the disk mid-plane is essential in order to provide accretion rates that agree with observational constraints.  

\acknowledgments

We thank Andrew Youdin, Sean O'Neill, and Matt Kunz for useful discussions and
suggestions regarding this work.  JBS, PJA, and KB acknowledge support from NASA through grants NNX09AB90G, NNX11AE12G and NNX13AI58G.  KB also acknowledges funding support from HST grant HST-AR-12814.03-A and from Tech-X Corp., Boulder, CO. XNB and JMS acknowledge support from the National Science Foundation through grant AST-0908269. PJA acknowledges support from NASA under grant HST-AR-12814 awarded by the Space Telescope Science Institute. XNB acknowledges support from program number HST-HF-51301.01-A provided by NASA through a Hubble Fellowship grant from the Space Telescope Science Institute, which is operated by the Association of Universities for Research in Astronomy, Inc., for NASA, under contact NAS 5-26555. 
This research was supported by an allocation of advanced computing resources provided by the National Science Foundation. The computations were performed on Kraken and Nautilus at the National Institute for Computational Sciences through XSEDE grant TG-AST120062.
This work also utilized the Janus supercomputer, which is supported by the National Science Foundation (award number CNS-0821794) and the University of Colorado Boulder. The Janus supercomputer is a joint effort of the University of Colorado Boulder, the University of Colorado Denver, and the National Center for Atmospheric Research.

\end{document}